\documentclass[notitlepage,superscriptaddress,aps,prl,twocolumn,noeprint,longbibliography]{revtex4-2}
\usepackage{amsmath}
\usepackage{amssymb}
\usepackage[sort&compress]{natbib}
\setcitestyle{numbers,comma,square}
\usepackage{upgreek}
\usepackage{mathtools}
\usepackage{graphicx}
\usepackage{afterpage}
\usepackage{amssymb}
\usepackage{epstopdf}
\usepackage[colorlinks=true,citecolor=blue,linkcolor=magenta]{hyperref}
\usepackage{tikz}
\usepackage{color}
\usepackage{bm}
\usepackage[caption=false,position=top,singlelinecheck=off,justification=raggedright]{subfig}

\DeclarePairedDelimiterX\braket[2]{\langle}{\rangle}{#1 \delimsize\vert #2}

\begin{document}
\title{ Nonequilibrium  Transport in a Superfluid Josephson Junction Chain: Is There Negative Differential Conductivity?}

\author{Samuel E. Begg}
\email{samuel.begg@apctp.org} 
\affiliation{Asia Pacific Center for Theoretical Physics, Pohang 37673, Korea}
\affiliation{Australian Research Council Centre of Excellence in Future Low-Energy Electronics Technologies, School of Mathematics and Physics, University of Queensland, St Lucia, Queensland 4072, Australia.}
\author{Matthew J. Davis}
\email{mdavis@uq.edu.au}
\affiliation{Australian Research Council Centre of Excellence in Future Low-Energy Electronics Technologies, School of Mathematics and Physics, University of Queensland, St Lucia, Queensland 4072, Australia.}\author{Matthew T. Reeves}
\email{m.reeves@uq.edu.au}
\affiliation{Australian Research Council Centre of Excellence in Future Low-Energy Electronics Technologies, School of Mathematics and Physics, University of Queensland, St Lucia, Queensland 4072, Australia.}

\date{\today}

\begin{abstract}
We consider the far-from-equilibrium quantum transport dynamics in a 1D Josephson junction chain of multi-mode Bose-Einstein condensates. We develop a theoretical model to examine the experiment of R.~Labouvie \textit{et al.} [Phys.~Rev.~Lett.~\textbf{115}, 050601 (2015)], wherein the phenomenon of negative differential conductivity (NDC) was reported in the refilling dynamics of an initially depleted site within the chain. We demonstrate that a unitary c-field description can quantitatively reproduce the experimental results over the full range of tunnel couplings, and requires no fitted parameters. With a view toward atomtronic implementations, we further demonstrate that the filling is strongly dependent on spatial phase variations stemming from quantum fluctuations. Our findings suggest that the interpretation of the device in terms of NDC is invalid outside of the weak coupling regime. Within this restricted regime, the device exhibits a hybrid behaviour of NDC and the AC Josephson effect. A simplified circuit model of the device will require an approach tailored to atomtronics that incorporates quantum fluctuations. 
\end{abstract}
\maketitle

\emph{Introduction ---}  Experiments in ultracold atomic gases allow unprecedented control over the dynamics of quantum systems. Developments such as arbitrary potential generation~\cite{Gauthier2016} and single-site addressing schemes~\cite{Bakr2009,Wurtz2009,Edge2015}, have driven interest in the development of so-called ``atomtronic" systems, consisting of  circuits and devices that leverage quantum features such as coherence or superfluid transport~\cite{Seaman2007,Eckel2014,Olsen2015,Kordas2015,Caliga2017,Burchianti2018,Ryu2020,Amico2021,Pandey2021,Perez2022,Grun2022}. Atomtronic simulators have also proven a powerful platform for probing a diverse range of nonequilibrium phenomena in quantum many-body systems within a simplified setting, for example quench dynamics~\cite{Kinoshita2006,Hung2013,Pigneur2018,Chen2019}, universal scaling phenomena~\cite{Lamporesi2013,Krinner2015,Erne2018,Eigen2018}, and quantum turbulence~\cite{navon2016emergence,navon2019synthetic,reeves2022turbulent}.

\begin{figure}[t]
\includegraphics[width=\columnwidth]{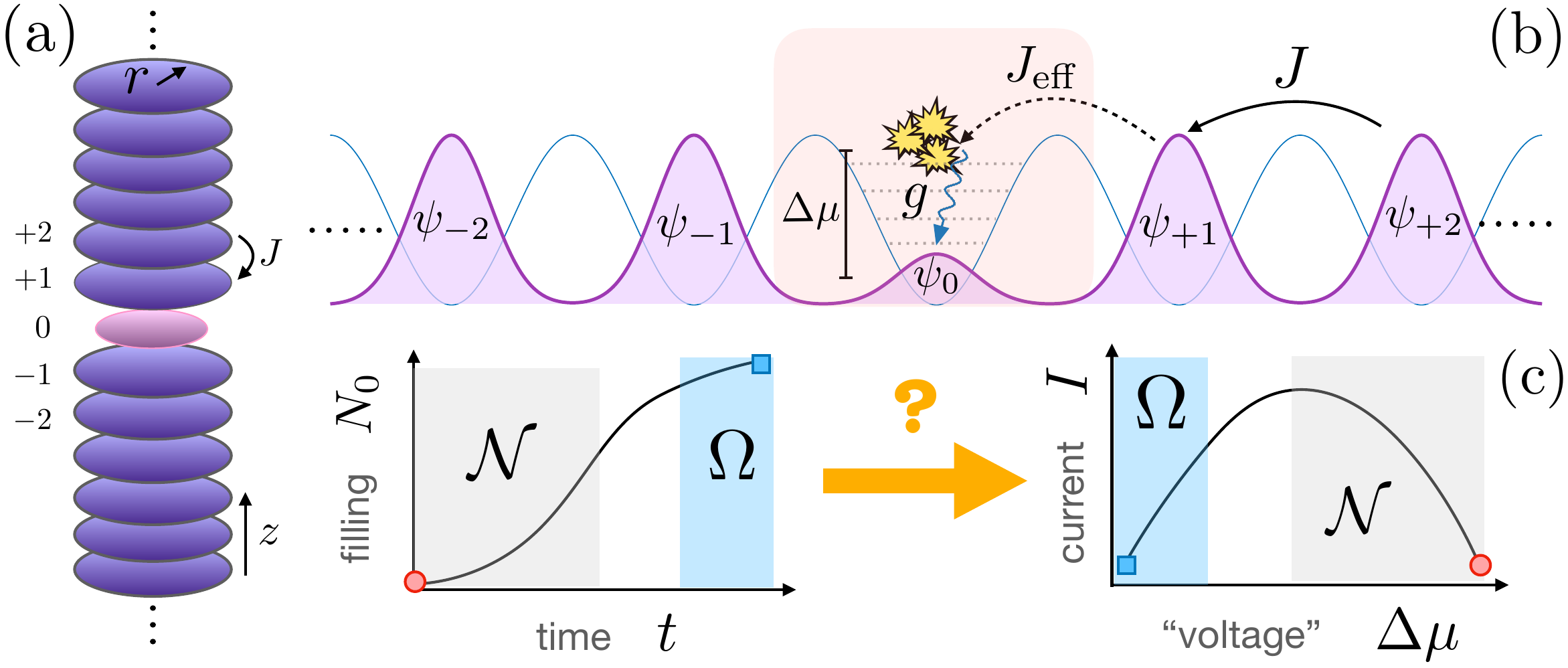}
\caption{(a) The multimode Josephson chain, considered experimentally in Ref.~\cite{Labouvie2015}, consists of an array of pancake-shaped condensates. (b) The effective hopping rate, $J_{\rm eff}$, onto the central depleted site is dependent on the atom number $N_0$, due to radial excitations; particles tunnel into radially excited modes and relax through the two-body interactions of strength $g$. (c) The atom number vs. time (left panel) was found to exhibit a characteristic ``S" shape. The inferred current-voltage relation (right panel) exhibited ohmic behaviour ($\Omega$) at small biases, and negative differential conductivity  ($\mathcal{N}$) at large biases.}
\label{fig:Schematic}
\end{figure}

Atomtronic  systems  utilizing  optical  lattice  potentials are  particularly suited for simulating nonequilibrium transport phenomena as occurs, for example, in the dynamics of Josephson junctions~\cite{Burchianti2018,Labouvie2015,Labouvie2016,Ceulemans2023}.
In a recent experiment, Labouvie \textit{et al.}~\cite{Labouvie2015} studied a ``multimode" Josephson array, formed by loading a large prolate atomic BEC into a 1D optical lattice. The resulting Josephson array,  shown schematically in Fig.~\ref{fig:Schematic}(a),  consisted of a chain of quasi-2D Bose-Einstein condensates (BECs), harmonically confined radially in $r$, and coupled by nearest-neighbour hopping $J$ along $z$. Using a site-selective electron beam, the experiment removed the atoms from the central site, and then observed the subsequent refilling dynamics  [Fig.~\ref{fig:Schematic}(b)].

The main finding of Ref.~\cite{Labouvie2015} was the observation of a novel current-filling relation in the central site.  
Unlike the AC oscillations characteristic of a standard (two-mode) Josephson junction, the system exhibited an emergent DC current; as shown schematically in Fig.~\ref{fig:Schematic}(c), the atom current 
onto the central site, $I = dN_0/dt$, was ohmic at small biases ($\Omega$). Meanwhile, at large biases ($\mathcal{N}$), the junction exhibited the phenomenon of negative differential conductivity (NDC), wherein the current decreased with increasing ``chemical voltage" $\Delta \mu$:
\begin{equation}
d I/d(\Delta \mu) < 0 \quad\quad \mathrm{(NDC)}. \label{eq:NDC}
\end{equation}
The emergent NDC was shown to be enabled by an atom-number-dependent effective tunnelling rate, $J_{\rm{eff}}$, arising due to many radially excited modes altering the tunnelling dynamics~[Fig.~\ref{fig:Schematic}(b)]. 

NDC is widely utilized in traditional semiconductor electronics~\cite{Shaw1992_1}, and has more recently been observed in systems  such as molecular break junctions~\cite{Perrin2014}, and multi-layer graphene~\cite{Britnell2013}. 
The strongly non-ohmic behaviour of NDC thus presents interesting prospects for atomtronic systems;  its characteristic multi-valued current-voltage relation 
[cf.~Fig.~\ref{fig:Schematic}(c)] 
enables the construction of elements such as diodes, thyristors and transistors~\cite{Shaw1992_1}. Furthermore, the mechanisms leading to NDC are typically nonequilibrium (``hot-carrier") phenomena~\cite{Volkov1969,Pamplin1970}, and its observation in an atomtronic setting thus hints that additional far-from-equilibrium quantum transport scenarios may be realizable with
ultracold atoms. This, in turn, might provide new insights into nonequilibrium quantum transport more generally, as ultracold atom systems offer a means to study phenomena difficult to probe in detail in the solid state~\cite{Brantut2013,Chien2015}. While the average filling behaviour 
observed experimentally 
has been reproduced by effective few-mode models~\cite{Labouvie2015,Mink2022}, a more complete model of the many-body dynamics presents a considerable theoretical challenge, and, to date, this has precluded a comprehensive understanding of the atomtronic NDC mechanism.

 In this Letter we develop a numerically tractable model
 for the nonequilibrium transport dynamics in the multimode Josephson  chain. In contrast to previous effective single-particle or few-mode approximations \cite{Labouvie2015,Olsen2016,Fischer2017,Mink2022}, we build a model within the framework of classical field theory which fully captures the many-body and multimode nature of the problem. Not only does our approach yield quantitative agreement with the experimental obervations (without any fitted parameters), it further allows the underlying mechanisms of the reported NDC to be probed in detail. We are able to demonstrate that $i)$: the device current is in fact nowhere ohmic, and that this apparent behaviour is due to the probabilistic nature of the  refilling seeded by quantum fluctuations; $ii)$: the interpretation of the current relation in terms of NDC is severely limited by the importance of phase dynamics and the far-from equilibrium nature of the system; and $iii)$ the device displays a combination of NDC- and AC Josephson-type behaviour. An important implication of our results for atomtronics is that electronic analogues should be approached with caution, and that a tailored theoretical approach specific to atomtronic junctions is likely required.

\emph{Model --- }
We model the multimode junction using c-field theory, also known as the truncated Wigner approximation (TWA) \cite{Steel1998,Polkovnikov2003a,Blakie2008}. Essentially, the TWA approximates the dynamics of the quantum system as a classical field (c-field), with quantum fluctuations accounted for at first order by adding a specified amount of noise to the initial conditions (for a review, see, e.g. Ref.~\cite{Blakie2008}).

We describe the system on each site $i$ using a stochastic classical field $\psi_i(\bm{x},t)$. The fields evolve according to coupled Gross-Pitaevskii equations
\begin{align}
i \hbar \partial_t \psi_i = \mathcal{L}\psi_i + J (\psi_{i-1}  + \psi_{i+1}), \label{eq:gpe}
\end{align}
where 
\begin{align}
\mathcal{L}\psi_i = \mathcal{P}\left\{\left[- \frac{\hbar \nabla^2}{2m} +V(\bm{x}) + g_2 |\psi_i|^2   \right]\psi_i \right\} ,
\label{eqn:GPEoperator}
\end{align}
is the (projected) Gross-Pitaevski operator and 
$V(\bm{x}) = \frac{1}{2} m \omega_r^2 (x^2+y^2)$
provides harmonic confinement in the $x$-$y$ plane with frequency $\omega_r.$
The interaction strength is $g_2 = g \int dz |w(z)|^4 $, where  $w(z)$ is the Wannier function associated with the optical lattice in the $z$-direction and $g = 4\pi \hslash^2 a_s/m$, with s-wave scattering length $a_s$. The tunnelling rate $J$ is determined by diagonalizing the optical lattice Hamiltonian (in the $z$-direction),
$\hat{H} = - \hslash^2 \partial_z^2/2m + V_0 \sin^2(2 \pi z / \lambda)$, and  using a tight-binding approximation (see Supplemental Material~\cite{SM}).  The projector $\mathcal{P}$ limits the dynamics of Eq.~(\ref{eq:gpe}) to the single particle modes that are highly occupied, (i.e., the classical field region) via an energy cutoff~\cite{SM}. Provided this is appropriately chosen, the model is insensitive to its precise value; we emphasise this leaves no free parameters in the model.

Tunnelling between neighbouring sites $i$ and $j=i\pm1$ depends on the overlap between the respective fields. This is evidenced through the evolution of the atom number on site $i$, $N_i$, 
\begin{align}
\frac{dN_i}{dt} = \frac{2J}{\hbar} \left(\eta_{i+1,i} \sqrt{N_{i+1}N_i} + \eta_{i-1,i} \sqrt{N_{i-1}N_i}\right), \label{eq:fillrate}
\end{align}
 with the ``Frank-Condon factors"
\begin{align}
\eta_{i,j}(t) &= {\text{Im}[\braket{\psi_{i}} {\psi_j}]}/{\sqrt{N_i N_{j}}} ,\nonumber\\ & = \frac{1}{\sqrt{N_iN_j}}\int d^2\bm{x} \sqrt{n_i(\bm{x}) n_j(\bm{x})} \sin [\varphi_{ij}(\bm{x})] ,
\label{eqn:FrankCondon}
\end{align} 
where $\psi_i(\bm{x}) = \sqrt{n_i(\bm{x})}e^{i \phi_i(\bm{x})}$ and $\varphi_{ij}(\bm{x}) \equiv \phi_i(\bm{x}) - \phi_j(\bm{x})$. The number-dependent filling mechanism is thus explicitly contained within the c-field formalism; this contrasts with the effective single particle description employed in Ref.~\cite{Labouvie2015}, where it was assumed to be $\eta_{i,j} \sim | \langle \psi_i | \psi_j \rangle |$, and approximated \emph{a posteriori} to be 
 linear in the particle difference $|N_i - N_j|$.

\begin{figure}[t]
\includegraphics[width=\columnwidth]{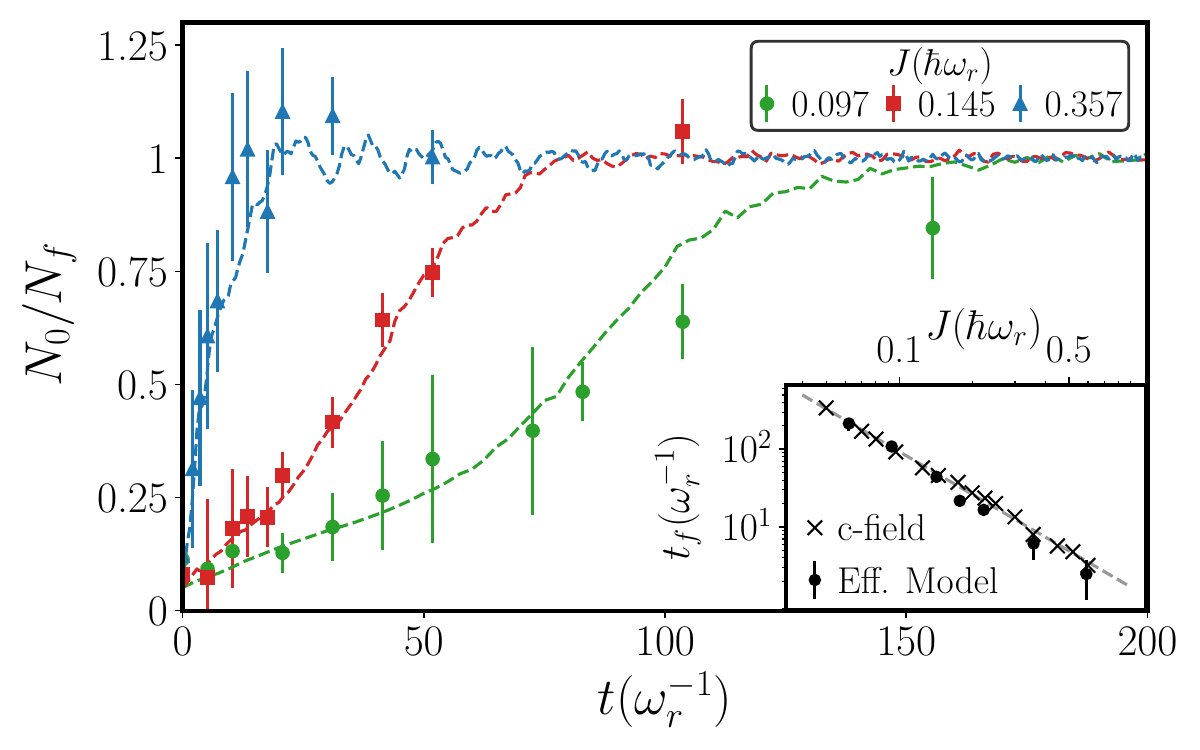}
\caption{Atom number $N_0/N_f$ vs time for different tunnel couplings $J$, comparing results obtained via 100 trajectories of the c-field model (\ref{eq:gpe}) (dashed lines) with the experimental data of Ref.~\cite{Labouvie2015} (points). Inset: Log-log plot of fill times $t_f$ vs $J$ as predicted by the c-field model (crosses) and the fitted effective single particle model presented in  Ref.~\cite{Labouvie2015} (dots). Dashed line shows a fit to the simulation data $t_f \propto J^{-\alpha}$ with $\alpha = 1.82(3)$. }
\label{fig:fullns}
\end{figure}

\emph{Results --- }  We solve Eq.~(\ref{eq:gpe}) via a pseudospectral method with quadrature using XMDS2~\cite{xmds2}, enabling accurate simulation of a large number of sites. Throughout we work in radial harmonic oscillator units, giving units for energy $\hbar \omega_r$, length $l_r = \sqrt{\hbar / m \omega_r}$, and  time $\omega_r^{-1}$.  We model the experiment using a uniform 21-site chain;  each site (for $i\neq0$)  contains $N_f\sim 700$ atoms~\cite{Labouvie2015}, and $g_2 \sim 0.2 \hbar \omega l_r^2$, giving $\mu_R \sim 7 \hbar \omega_r$.  
As per Ref.~\cite{Labouvie2015}, the central site ($i=0$) is populated such that the atom number is $N_0/N_f \sim 5\%$. From a detailed modelling of the experimental preparation protocol, we find that, at the end of the preparation sequence, the left and right chains become nearly uncorrelated in their phases. For the initial conditions, we therefore multiply one of the chains by a random phase to mimic this phase diffusion.
Full details of the experimental modelling setup and simulation method are provided in the Supplemental Material~\cite{SM}.

We first compare the results of the numerical simulations directly against the experimental observations of Ref.~\cite{Labouvie2015}. In Fig.~\ref{fig:fullns} we show the atom number $N_0/N_f$ vs. time  for three different tunnelling couplings $J$. The c-field model is in excellent agreement across the full range of coupling values. The model produces the characteristic ``S"-shaped filling curves, which suggest ohmic behaviour at small voltages (late times) and NDC at large voltages (early times) [cf. Fig.~\ref{fig:Schematic}(c)]. Figure \ref{fig:fullns} (inset) shows the filling time, $t_f$, defined as the time at which $N_0$ reaches $2/3$ of its final value~\cite{Labouvie2015}. We find the filling time obeys the power-law $t_f \propto J^{-\alpha}$, with  $\alpha = 1.82(3)$. This is in good agreement with the value of $\alpha = 1.9(1)$ obtained in Ref.~\cite{Labouvie2015} which utilized an effective single particle model with a fitted decoherence parameter.

\begin{figure}[t]
\subfloat{\includegraphics[width=4.3cm] {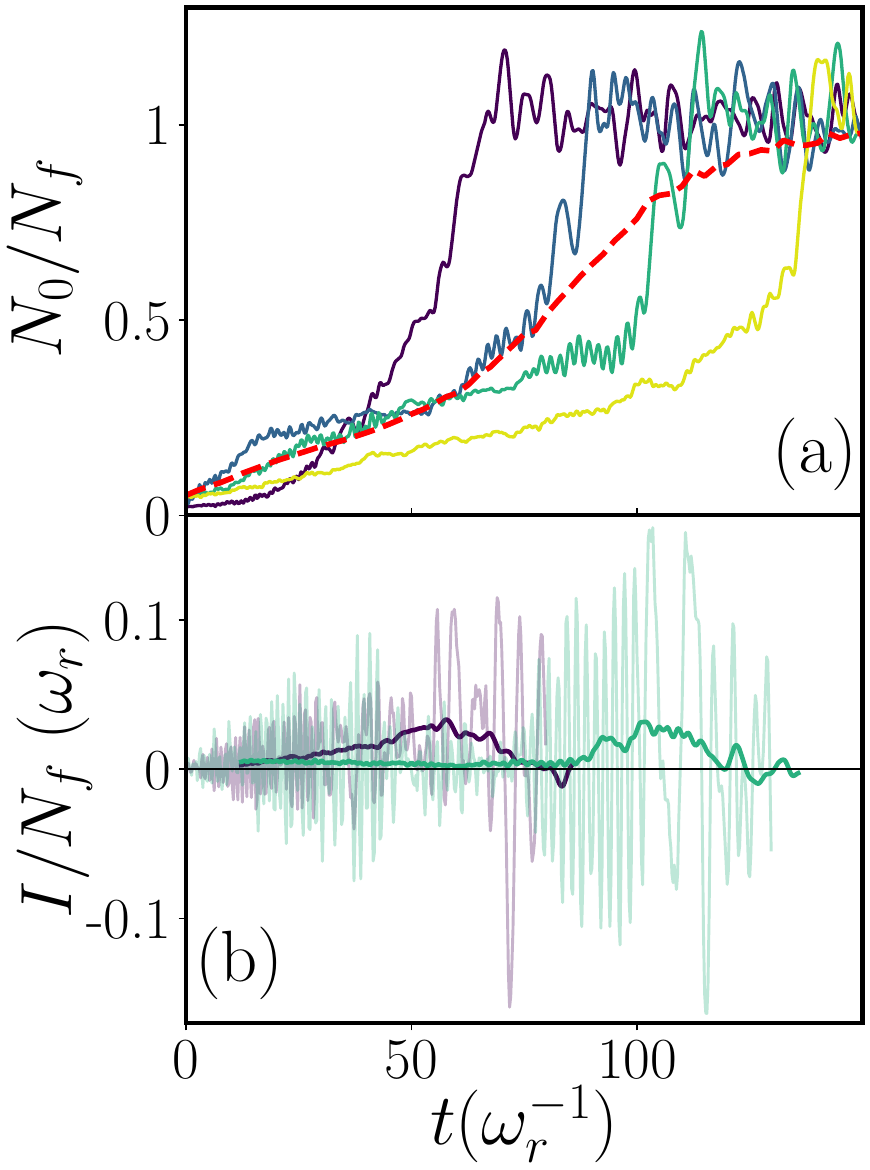}\label{fig:NDC}} 
\subfloat{\hspace{0.02cm} \includegraphics[width=4.25cm]{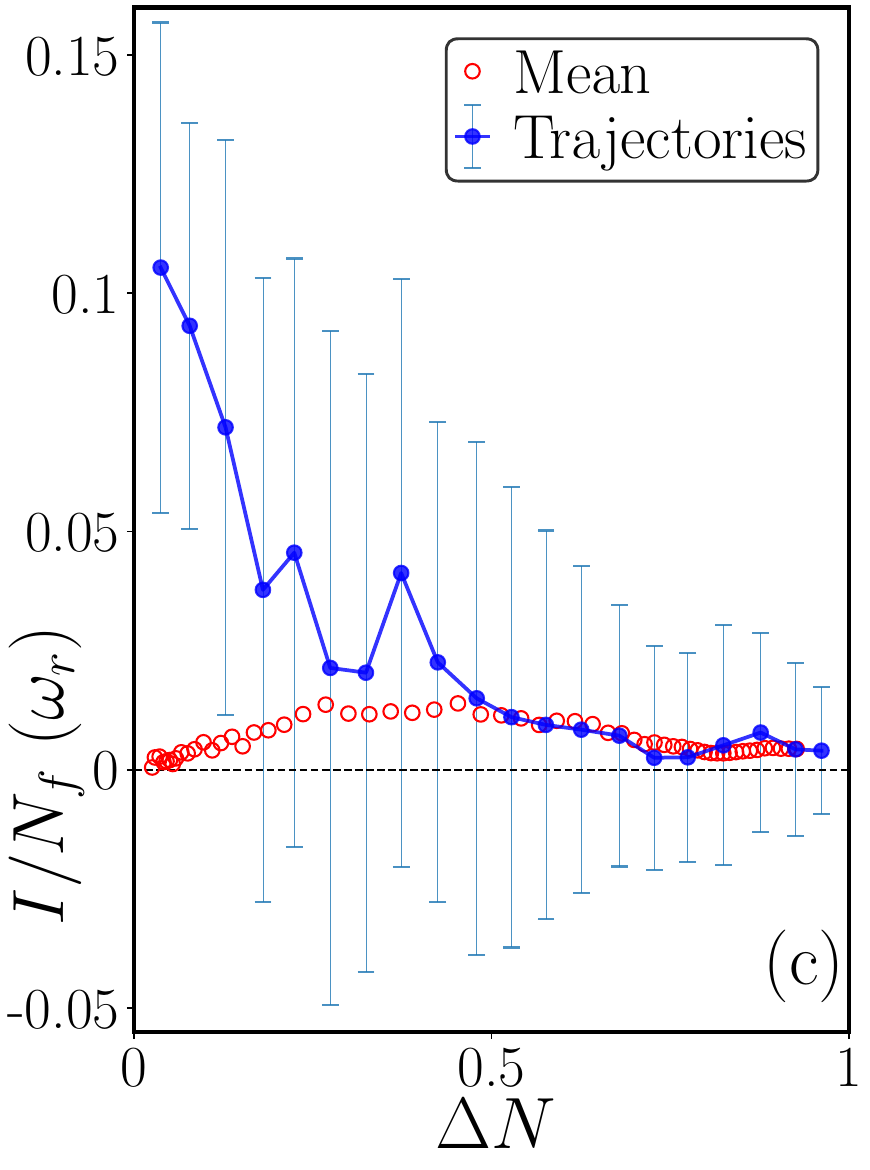}\label{fig:currentpotential}}
\caption{(a) Comparison of the atom number $N_0/N_f$ vs time for five individual trajectories (solid lines), compared with the mean of 100 trajectories (dashed line) with $J/\hbar \omega_r = 0.097 $~\protect\cite{normalisation_of_trajectories}. (b) Normalized particle current into the central site $I$ vs time for two trajectories (light lines). The dark lines show the corresponding rolling averages. (c) $I$ vs atom number difference (normalized) $\Delta N = (N_f-N_0)/N_f$, comparing the rate of change of the mean atom number $I_m = d\langle N_0 \rangle/dt$ (circles), against the  average trajectory current $I_t= \langle dN_0/dt \rangle$ (dots). Error bars show the standard deviation~\cite{Fig3footnote}. }
\label{fig:mean_vs_trajectories}
\end{figure}

Having established that the model quantitatively reproduces experimental observations, we now turn to gaining deeper insight into the mechanisms underlying the NDC-type behaviour.The experimental measurement protocol of Labouvie \emph{et al.}~\cite{Labouvie2015} was destructive, so that every data point in Fig.~\ref{fig:fullns}(a) is an average over several experimental runs beginning with an independent condensate.  Consequently, the results of single experiments cannot be inferred. However, for the TWA, the individual simulation trajectories approximately correspond to the behaviour of single experimental runs~\cite{Blakie2008}.  In Fig.~\ref{fig:mean_vs_trajectories}(a) we compare the mean atom number vs.~time, $\langle N_0(t) \rangle$  (red-dashed line), with $N_0(t)$ for several  individual simulation trajectories, with $J/\hbar\omega_r=0.097$~\protect\cite{normalisation_of_trajectories}.

%In Fig.~\ref{fig:mean_vs_trajectories}(a) we compare the mean atom number vs.~time, $\langle N_0(t) \rangle$  (red-dashed line), with, $N_0(t)$ for several  individual simulation trajectories, with $J/\hbar\omega_r=0.097$~\protect\cite{normalisation_of_trajectories}. Within the TWA, the trajectories approximately correspond to the behaviour in individual experimental runs~\cite{Blakie2008}. 
%This allows deeper insight than was available experimentally,  where, due to the destructive nature of the measurement protocol, every data point corresponds to a different condensate. Therefore, only the mean results were available and the behavior of individual trajectory dynamics could not be inferred in Ref.~\cite{Labouvie2015}. 

 In the simulations the individual trajectories and overall  curve for the mean $\langle N_0(t) \rangle$ exhibit similar behaviour at early times but markedly different behaviour at late times. In particular, the characteristic ``S"-shape  that was observed experimentally  [cf. Fig.~\ref{fig:fullns}] only appears in the mean  of the simulation data; it emerges due to the probabilistic filling of the individual trajectories, which have different filling times due to quantum noise. In individual trajectories [Fig.~\ref{fig:mean_vs_trajectories}(a)], the current can be seen to in fact \emph{accelerate} rather than decelerate during the latter stages of filling.  Unlike the smooth growth of the mean, the current in individual trajectories is a rapidly oscillating chirped signal in time [Fig.~\ref{fig:mean_vs_trajectories}(b)].  In many trajectories a rapid growth of the atom number occurs during latter stages of the filling [Fig.~\ref{fig:mean_vs_trajectories}(a)]; this is accompanied by oscillations that increase in amplitude and decrease in frequency as time progresses. This rapid growth phase also coincides with a sudden growth in the condensate fraction~\cite{SM}; the late-stage oscillations are thus strongly suggestive of the ``AC Josephson effect"  previously observed in, for example, Refs.~\cite{Pereverzev1997,Levy2007}.

Differences between the mean and individual trajectories are further highlighted in Fig.~\ref{fig:mean_vs_trajectories}(c), which shows the rate of change of the mean atom number $I_m = d\langle N_0 \rangle/dt$ and the average trajectory current $I_t = \langle dN_0/dt \rangle$, plotted against the atom number difference $\Delta N = (N_f-N_0)/N_f$.
For $\Delta N \gtrsim 0.6$ ($t\lesssim t_f$), we find $I_t \approx I_m$, with both exhibiting the NDC characteristic Eq.~(\ref{eq:NDC}). However, for  $\Delta N \lesssim 0.6$ ($t \gtrsim t_f$), we find $I_t\neq I_m$. Notably, the ohmic region ($I \propto \Delta N$) seen in $I_m$ is only apparent in the average. Contrarily, the trajectory data in Fig.~\ref{fig:mean_vs_trajectories}(c) suggest that the individual trajectories instead exhibit the NDC characteristic over the entire range of $\Delta N$. The rapid current oscillations in the trajectories [Fig.~\ref{fig:mean_vs_trajectories}(b)] also result in significant fluctuations in $I_t$ [Fig.~\ref{fig:mean_vs_trajectories}(c), errorbars]; these are in fact larger than the average currents for most values of $\Delta N$. Note that we have refrained from defining a ``chemical voltage" $\Delta \mu$ as in Ref.~\cite{Labouvie2015}, because this requires the assumption of quasi-equilibrium, whereas $\Delta N$ is always defined. To support this, in the Supplemental Material \cite{SM} we demonstrate that the experiment~\cite{Labouvie2015} is far-from equilibrium.

The results in Fig.~\ref{fig:mean_vs_trajectories}(a) show that the trajectories certainly exhibit the NDC characteristic Eq.~(\ref{eq:NDC}). However, this alone is not sufficient for the behaviour to be classified as NDC; it is also necessary to consider the role played by the relative phases between sites. NDC in the usual sense would require that the phase information is unimportant, such that the current can be characterized simply through the particle number difference as  $I = I(\Delta N)$. By contrast, Josephson junctions are characterized by both an imbalance (voltage) and their intrinsic phase difference, i.e., $I = I(\Delta N, \varphi)$, and for this reason they are not readily classified as NDC elements despite exhibiting quasi-NDC-type behaviour~\cite{Shaw1992_1,Shaw1992_6}.

\begin{figure}[t!]
\subfloat{\includegraphics[width=\columnwidth]{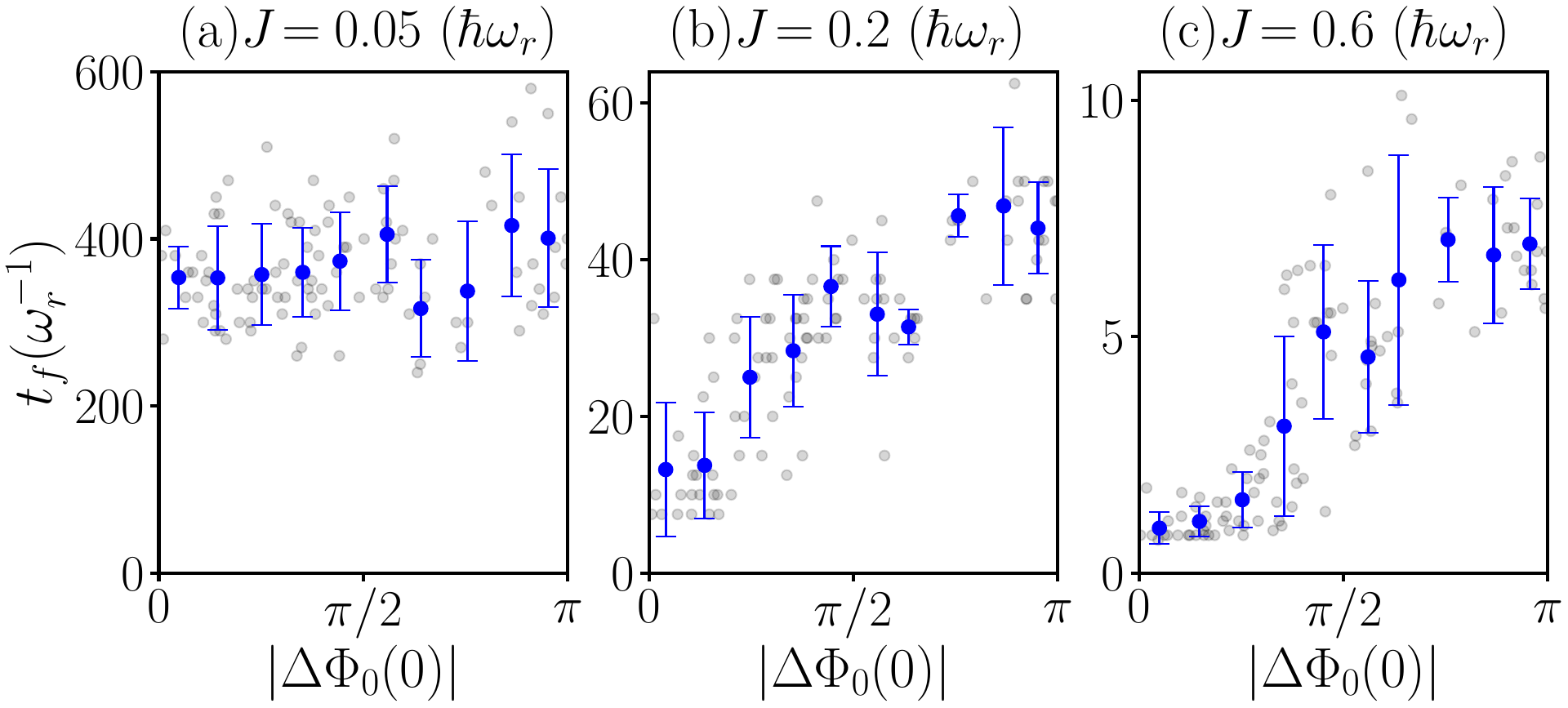}\label{fig:dph_vs_phase}}
\caption{Influence of the initial phase on the filling dynamics. (a) Filling time $t_f$ vs initial phase difference $|\Delta \Phi_0(0)|$ for different tunnelling strengths $J$. Light markers show individual trajectories, dark markers with errorbars show average and standard deviation (obtained by histogramming over 10 bins in the range $|\Delta \Phi_0| \in [0,\pi])$. 
}
\label{fig:relativePhase}
\end{figure}

To investigate the role of the phase in the dynamics, we consider the phase difference between the nearest-neighbour sites $i = \pm 1$ in individual trajectories. We define $\Delta \Phi_i = \Phi_{i+ 1} - \Phi_{i-1}$, where $\Phi_i$ is the spatially averaged phase $\Phi_i = \mathcal{A}^{-1} \int_\mathcal{A} d^2 \bm{x} \; \phi_i \equiv \langle \phi_i \rangle_{\bm{x}} $ (where $\mathcal{A}$ is a central region in well $i$).  Figure \ref{fig:relativePhase} shows the dependence of the filling dynamics on the initial phase difference across the central site, $\Delta\Phi_0$, for three different values of $J$. For small couplings ($J/\hbar \omega_r= 0.05$), the filling time is found to be approximately independent of the initial phase difference. However, for moderate tunnelling $(J/\hbar \omega_r = 0.2)$, when the left and right chains are initially out of phase $(|\Delta \Phi_0| \sim \pi)$ the filling time is approximately double that of the synchronized case $|\Delta \Phi_0| \sim 0 $. For large couplings ($J/\hbar \omega_r = 0.6$), this discrepancy becomes even larger, yielding a factor of $\sim$ 6 difference in the filling times between the zero-phase and $\pi$-phase difference scenarios. For $|\Delta\Phi_0| \sim \pi$, we also observe a significant increase in the fluctuations of the filling time, as seen in the large standard deviations indicated by the error bars. 

\emph{Discussion --- }  Our results suggest that the NDC interpretation put forward in Ref.~\cite{Labouvie2015} is only valid at small values of $J$ and during the early stages of the filling dynamics, where phase coherence is unimportant.  For most regimes the phase difference is a crucial element in the dynamics.  Indeed, Eq.~(\ref{eq:fillrate}) shows that both the particle number difference and phase differences drive the current, since $\eta_{i,j} \propto \Im \{\langle \psi_i | \psi_j \rangle\}$ rather than $|\langle \psi_i|\psi_j \rangle |$ as was assumed in Ref.~\cite{Labouvie2015}.  To consider the average contributions from the phase, one should consider the  
the conjugate equation to Eq.~(\ref{eq:fillrate}), which governs the evolution of the spatially-averaged phase $\Phi_j$:
\begin{equation}
\hbar \partial_t \Phi_j + \mu^{\rm eff}_{j} =  J \zeta_j,
\label{eqn:Phi}
\end{equation}
with
\begin{equation}
    \mu^{\rm eff}_{j} =  \left\langle \tfrac{1}{2} m \mathbf{u}_j^2 + g_2 n_j + V - \tfrac{\hbar^2}{2 m} \tfrac{\nabla^2 \sqrt{n_j}}{\sqrt{n_j}} \right\rangle_{\bm{x}},
\end{equation}
where $\mathbf{u}_j(\bm{x},t) =  \hbar\nabla \phi_j(\bm{x},t)/m$ is the velocity field on site $j$, and
\begin{equation}
    \zeta_j = \left\langle \sqrt{\tfrac{n_{j+1}}{n_j}} \cos( \varphi_{j+1,j}) + \sqrt{\tfrac{n_{j-1}}{n_j}} \cos( \varphi_{j-1,j} ) \right\rangle_{\bm{x}}.
    \label{eqn:zeta}
\end{equation}
The evolution of the average variables $(N_j,\Phi_j)$ relies on a rather non-trivial coarse graining over the internal dynamics of the sites; this depends on the instantaneous density fields $n_j$, relative phases $\varphi_{ij}$, and on-site velocity fields $\mathbf{u}_j$. 

Our model clearly reproduces the observed experimental behavior, and thus serves as a natural starting point for developing a reduced ``lumped-element"-type description. However, whether the full dynamics of Eq.~(\ref{eq:gpe}) can indeed be reduced to a simple effective model in terms of $(N_j,\Phi_j)$ remains an  open problem. We anticipate the results could be significantly affected by, e.g., thermal effects (condensate fraction), collective modes, or quantized vortices, which all affect the average phase coherence. While a comprehensive study of finite temperature effects is beyond the scope of the present work, we performed simulations of a mostly incoherent process ($<5\%$ condensate fraction in the reservoirs), which further support the interpretations put forward here (see~\cite{SM}). In particular, the acceleration of the current in individual trajectories is absent for the incoherent filling process, indicating that phase coherence is essential for this process to occur. 
Finally, we note that many of the system's qualitative dynamical features can be reproduced by a simple ``coherent reservoir" model, wherein the reservoir is treated as a coherent AC drive.
This model was described in Ref.~\cite{Reeves2021} to model a related experiment wherein dissipation was introduced to the central site~\cite{Labouvie2016}, and may serve as a useful starting point for a simplified description of multimode Josephson junction dynamics.

\emph{Conclusions --- } We have characterized the nonequilibrium dynamics of a many-body, multimode bosonic Josephson chain, and demonstrated that a c-field description can quantitatively reproduce the observations of Ref.~\cite{Labouvie2015} with no fitted parameters. Our model suggests that the dynamics are strongly dependent on both the atom number and phase. We argue this makes the NDC interpretation invalid for the majority of the couplings considered in Ref.~\cite{Labouvie2015}. The filling appears to be a ``two-stage" process: our results do suggest that the NDC interpretation is valid at small $J$ and during the initial (gradual)  phase of the filling, before phase coherence is established. However, this interpretation is complicated in latter (rapid) stages of the filling by the emergence of phase coherence and strong Josephson oscillations. Additionally, significant (quantum) fluctuations are present, and will likely need to be incorporated into any simplified effective model that quantitatively reproduces the salient features of the dynamics. 
From a device perspective, it would be worthwhile considering how the atomtronic junction behaves under an enforced chemical potential bias, rather than under free dynamics, so as to be considered as a single component of a larger DC or AC atomtronic circuit. 

\emph{Acknowledgements --- } We acknowledge useful discussions with H.~Ott and A.~S.~Bradley. This research was partially supported by the Australian Research Council Centre of Excellence in Future Low-Energy Electronics Technologies (FLEET, Project No.~CE170100039) and funded by the Australian Government. M.T.R.~is supported by an Australian Research Council Discovery Early Career Researcher Award (DECRA), Project No. DE220101548. S.E.B.~acknowledges the support of the Young Scientist Training Program at the Asia Pacific Center for Theoretical Physics.

\bibliographystyle{apsrev4-2}

\begin{thebibliography}{55}%
\makeatletter
\providecommand \@ifxundefined [1]{%
 \@ifx{#1\undefined}
}%
\providecommand \@ifnum [1]{%
 \ifnum #1\expandafter \@firstoftwo
 \else \expandafter \@secondoftwo
 \fi
}%
\providecommand \@ifx [1]{%
 \ifx #1\expandafter \@firstoftwo
 \else \expandafter \@secondoftwo
 \fi
}%
\providecommand \natexlab [1]{#1}%
\providecommand \enquote  [1]{``#1''}%
\providecommand \bibnamefont  [1]{#1}%
\providecommand \bibfnamefont [1]{#1}%
\providecommand \citenamefont [1]{#1}%
\providecommand \href@noop [0]{\@secondoftwo}%
\providecommand \href [0]{\begingroup \@sanitize@url \@href}%
\providecommand \@href[1]{\@@startlink{#1}\@@href}%
\providecommand \@@href[1]{\endgroup#1\@@endlink}%
\providecommand \@sanitize@url [0]{\catcode `\\12\catcode `\$12\catcode
  `\&12\catcode `\#12\catcode `\^12\catcode `\_12\catcode `\%12\relax}%
\providecommand \@@startlink[1]{}%
\providecommand \@@endlink[0]{}%
\providecommand \url  [0]{\begingroup\@sanitize@url \@url }%
\providecommand \@url [1]{\endgroup\@href {#1}{\urlprefix }}%
\providecommand \urlprefix  [0]{URL }%
\providecommand \Eprint [0]{\href }%
\providecommand \doibase [0]{https://doi.org/}%
\providecommand \selectlanguage [0]{\@gobble}%
\providecommand \bibinfo  [0]{\@secondoftwo}%
\providecommand \bibfield  [0]{\@secondoftwo}%
\providecommand \translation [1]{[#1]}%
\providecommand \BibitemOpen [0]{}%
\providecommand \bibitemStop [0]{}%
\providecommand \bibitemNoStop [0]{.\EOS\space}%
\providecommand \EOS [0]{\spacefactor3000\relax}%
\providecommand \BibitemShut  [1]{\csname bibitem#1\endcsname}%
\let\auto@bib@innerbib\@empty
%</preamble>
\bibitem [{\citenamefont {Gauthier}\ \emph {et~al.}(2016)\citenamefont
  {Gauthier}, \citenamefont {Lenton}, \citenamefont {Parry}, \citenamefont
  {Baker}, \citenamefont {Davis}, \citenamefont {Rubinsztein-Dunlop},\ and\
  \citenamefont {Neely}}]{Gauthier2016}%
  \BibitemOpen
  \bibfield  {author} {\bibinfo {author} {\bibfnamefont {G.}~\bibnamefont
  {Gauthier}}, \bibinfo {author} {\bibfnamefont {I.}~\bibnamefont {Lenton}},
  \bibinfo {author} {\bibfnamefont {N.~M.}\ \bibnamefont {Parry}}, \bibinfo
  {author} {\bibfnamefont {M.}~\bibnamefont {Baker}}, \bibinfo {author}
  {\bibfnamefont {M.~J.}\ \bibnamefont {Davis}}, \bibinfo {author}
  {\bibfnamefont {H.}~\bibnamefont {Rubinsztein-Dunlop}},\ and\ \bibinfo
  {author} {\bibfnamefont {T.~W.}\ \bibnamefont {Neely}},\ }\href
  {https://opg.optica.org/optica/fulltext.cfm?uri=optica-3-10-1136&id=351043}
  {\bibfield  {journal} {\bibinfo  {journal} {Optica}\ }\textbf {\bibinfo
  {volume} {3}},\ \bibinfo {pages} {1136} (\bibinfo {year} {2016})}\BibitemShut
  {NoStop}%
\bibitem [{\citenamefont {Bakr}\ \emph {et~al.}(2009)\citenamefont {Bakr},
  \citenamefont {Gillen}, \citenamefont {Peng}, \citenamefont {F{\"o}lling},\
  and\ \citenamefont {Greiner}}]{Bakr2009}%
  \BibitemOpen
  \bibfield  {author} {\bibinfo {author} {\bibfnamefont {W.~S.}\ \bibnamefont
  {Bakr}}, \bibinfo {author} {\bibfnamefont {J.~I.}\ \bibnamefont {Gillen}},
  \bibinfo {author} {\bibfnamefont {A.}~\bibnamefont {Peng}}, \bibinfo {author}
  {\bibfnamefont {S.}~\bibnamefont {F{\"o}lling}},\ and\ \bibinfo {author}
  {\bibfnamefont {M.}~\bibnamefont {Greiner}},\ }\href
  {https://www.nature.com/articles/nature08482} {\bibfield  {journal} {\bibinfo
   {journal} {Nature}\ }\textbf {\bibinfo {volume} {462}},\ \bibinfo {pages}
  {74} (\bibinfo {year} {2009})}\BibitemShut {NoStop}%
\bibitem [{\citenamefont {W\"urtz}\ \emph {et~al.}(2009)\citenamefont
  {W\"urtz}, \citenamefont {Langen}, \citenamefont {Gericke}, \citenamefont
  {Koglbauer},\ and\ \citenamefont {Ott}}]{Wurtz2009}%
  \BibitemOpen
  \bibfield  {author} {\bibinfo {author} {\bibfnamefont {P.}~\bibnamefont
  {W\"urtz}}, \bibinfo {author} {\bibfnamefont {T.}~\bibnamefont {Langen}},
  \bibinfo {author} {\bibfnamefont {T.}~\bibnamefont {Gericke}}, \bibinfo
  {author} {\bibfnamefont {A.}~\bibnamefont {Koglbauer}},\ and\ \bibinfo
  {author} {\bibfnamefont {H.}~\bibnamefont {Ott}},\ }\href
  {https://doi.org/10.1103/PhysRevLett.103.080404} {\bibfield  {journal}
  {\bibinfo  {journal} {Phys. Rev. Lett.}\ }\textbf {\bibinfo {volume} {103}},\
  \bibinfo {pages} {080404} (\bibinfo {year} {2009})}\BibitemShut {NoStop}%
\bibitem [{\citenamefont {Edge}\ \emph {et~al.}(2015)\citenamefont {Edge},
  \citenamefont {Anderson}, \citenamefont {Jervis}, \citenamefont {McKay},
  \citenamefont {Day}, \citenamefont {Trotzky},\ and\ \citenamefont
  {Thywissen}}]{Edge2015}%
  \BibitemOpen
  \bibfield  {author} {\bibinfo {author} {\bibfnamefont {G.~J.~A.}\
  \bibnamefont {Edge}}, \bibinfo {author} {\bibfnamefont {R.}~\bibnamefont
  {Anderson}}, \bibinfo {author} {\bibfnamefont {D.}~\bibnamefont {Jervis}},
  \bibinfo {author} {\bibfnamefont {D.~C.}\ \bibnamefont {McKay}}, \bibinfo
  {author} {\bibfnamefont {R.}~\bibnamefont {Day}}, \bibinfo {author}
  {\bibfnamefont {S.}~\bibnamefont {Trotzky}},\ and\ \bibinfo {author}
  {\bibfnamefont {J.~H.}\ \bibnamefont {Thywissen}},\ }\href
  {https://link.aps.org/doi/10.1103/PhysRevA.92.063406} {\bibfield  {journal}
  {\bibinfo  {journal} {Phys. Rev. A}\ }\textbf {\bibinfo {volume} {92}},\
  \bibinfo {pages} {063406} (\bibinfo {year} {2015})}\BibitemShut {NoStop}%
\bibitem [{\citenamefont {Seaman}\ \emph {et~al.}(2007)\citenamefont {Seaman},
  \citenamefont {Kr{\"a}mer}, \citenamefont {Anderson},\ and\ \citenamefont
  {Holland}}]{Seaman2007}%
  \BibitemOpen
  \bibfield  {author} {\bibinfo {author} {\bibfnamefont {B.}~\bibnamefont
  {Seaman}}, \bibinfo {author} {\bibfnamefont {M.}~\bibnamefont {Kr{\"a}mer}},
  \bibinfo {author} {\bibfnamefont {D.}~\bibnamefont {Anderson}},\ and\
  \bibinfo {author} {\bibfnamefont {M.}~\bibnamefont {Holland}},\ }\href
  {https://journals.aps.org/pra/abstract/10.1103/PhysRevA.75.023615} {\bibfield
   {journal} {\bibinfo  {journal} {Phys. Rev. A}\ }\textbf {\bibinfo {volume}
  {75}},\ \bibinfo {pages} {023615} (\bibinfo {year} {2007})}\BibitemShut
  {NoStop}%
\bibitem [{\citenamefont {Eckel}\ \emph {et~al.}(2014)\citenamefont {Eckel},
  \citenamefont {Lee}, \citenamefont {Jendrzejewski}, \citenamefont {Murray},
  \citenamefont {Clark}, \citenamefont {Lobb}, \citenamefont {Phillips},
  \citenamefont {Edwards},\ and\ \citenamefont {Campbell}}]{Eckel2014}%
  \BibitemOpen
  \bibfield  {author} {\bibinfo {author} {\bibfnamefont {S.}~\bibnamefont
  {Eckel}}, \bibinfo {author} {\bibfnamefont {J.~G.}\ \bibnamefont {Lee}},
  \bibinfo {author} {\bibfnamefont {F.}~\bibnamefont {Jendrzejewski}}, \bibinfo
  {author} {\bibfnamefont {N.}~\bibnamefont {Murray}}, \bibinfo {author}
  {\bibfnamefont {C.~W.}\ \bibnamefont {Clark}}, \bibinfo {author}
  {\bibfnamefont {C.~J.}\ \bibnamefont {Lobb}}, \bibinfo {author}
  {\bibfnamefont {W.~D.}\ \bibnamefont {Phillips}}, \bibinfo {author}
  {\bibfnamefont {M.}~\bibnamefont {Edwards}},\ and\ \bibinfo {author}
  {\bibfnamefont {G.~K.}\ \bibnamefont {Campbell}},\ }\href
  {https://www.nature.com/articles/nature12958} {\bibfield  {journal} {\bibinfo
   {journal} {Nature}\ }\textbf {\bibinfo {volume} {506}},\ \bibinfo {pages}
  {200} (\bibinfo {year} {2014})}\BibitemShut {NoStop}%
\bibitem [{\citenamefont {Olsen}\ and\ \citenamefont
  {Bradley}(2015)}]{Olsen2015}%
  \BibitemOpen
  \bibfield  {author} {\bibinfo {author} {\bibfnamefont {M.~K.}\ \bibnamefont
  {Olsen}}\ and\ \bibinfo {author} {\bibfnamefont {A.~S.}\ \bibnamefont
  {Bradley}},\ }\href
  {https://journals.aps.org/pra/abstract/10.1103/PhysRevA.91.043635} {\bibfield
   {journal} {\bibinfo  {journal} {Phys. Rev. A}\ }\textbf {\bibinfo {volume}
  {91}},\ \bibinfo {pages} {043635} (\bibinfo {year} {2015})}\BibitemShut
  {NoStop}%
\bibitem [{\citenamefont {Kordas}\ \emph {et~al.}(2015)\citenamefont {Kordas},
  \citenamefont {Witthaut},\ and\ \citenamefont {Wimberger}}]{Kordas2015}%
  \BibitemOpen
  \bibfield  {author} {\bibinfo {author} {\bibfnamefont {G.}~\bibnamefont
  {Kordas}}, \bibinfo {author} {\bibfnamefont {D.}~\bibnamefont {Witthaut}},\
  and\ \bibinfo {author} {\bibfnamefont {S.}~\bibnamefont {Wimberger}},\ }\href
  {https://onlinelibrary.wiley.com/doi/full/10.1002/andp.201400189?casa_token=MZOZ_zpl3c8AAAAA%3ABAZnOTFuOhSjuwnPetu-Kj-bCcwMRkA1Tjo51Pq6ypxPApCMK6IH0yd8qhN0HKicF-STfwvrkxNT7AHP}
  {\bibfield  {journal} {\bibinfo  {journal} {Ann. Phys.}\ }\textbf {\bibinfo
  {volume} {527}},\ \bibinfo {pages} {619} (\bibinfo {year}
  {2015})}\BibitemShut {NoStop}%
\bibitem [{\citenamefont {Caliga}\ \emph {et~al.}(2017)\citenamefont {Caliga},
  \citenamefont {Straatsma},\ and\ \citenamefont {Anderson}}]{Caliga2017}%
  \BibitemOpen
  \bibfield  {author} {\bibinfo {author} {\bibfnamefont {S.~C.}\ \bibnamefont
  {Caliga}}, \bibinfo {author} {\bibfnamefont {C.~J.~E.}\ \bibnamefont
  {Straatsma}},\ and\ \bibinfo {author} {\bibfnamefont {D.~Z.}\ \bibnamefont
  {Anderson}},\ }\href
  {https://iopscience.iop.org/article/10.1088/1367-2630/aa56d8/meta} {\bibfield
   {journal} {\bibinfo  {journal} {New J. Phys.}\ }\textbf {\bibinfo {volume}
  {19}},\ \bibinfo {pages} {013036} (\bibinfo {year} {2017})}\BibitemShut
  {NoStop}%
\bibitem [{\citenamefont {Burchianti}\ \emph {et~al.}(2018)\citenamefont
  {Burchianti}, \citenamefont {Scazza}, \citenamefont {Amico}, \citenamefont
  {Valtolina}, \citenamefont {Seman}, \citenamefont {Fort}, \citenamefont
  {Zaccanti}, \citenamefont {Inguscio},\ and\ \citenamefont
  {Roati}}]{Burchianti2018}%
  \BibitemOpen
  \bibfield  {author} {\bibinfo {author} {\bibfnamefont {A.}~\bibnamefont
  {Burchianti}}, \bibinfo {author} {\bibfnamefont {F.}~\bibnamefont {Scazza}},
  \bibinfo {author} {\bibfnamefont {A.}~\bibnamefont {Amico}}, \bibinfo
  {author} {\bibfnamefont {G.}~\bibnamefont {Valtolina}}, \bibinfo {author}
  {\bibfnamefont {J.}~\bibnamefont {Seman}}, \bibinfo {author} {\bibfnamefont
  {C.}~\bibnamefont {Fort}}, \bibinfo {author} {\bibfnamefont {M.}~\bibnamefont
  {Zaccanti}}, \bibinfo {author} {\bibfnamefont {M.}~\bibnamefont {Inguscio}},\
  and\ \bibinfo {author} {\bibfnamefont {G.}~\bibnamefont {Roati}},\ }\href
  {https://link.aps.org/doi/10.1103/PhysRevLett.120.025302} {\bibfield
  {journal} {\bibinfo  {journal} {Phys. Rev. Lett.}\ }\textbf {\bibinfo
  {volume} {120}},\ \bibinfo {pages} {025302} (\bibinfo {year}
  {2018})}\BibitemShut {NoStop}%
\bibitem [{\citenamefont {Ryu}\ \emph {et~al.}(2020)\citenamefont {Ryu},
  \citenamefont {Samson},\ and\ \citenamefont {Boshier}}]{Ryu2020}%
  \BibitemOpen
  \bibfield  {author} {\bibinfo {author} {\bibfnamefont {C.}~\bibnamefont
  {Ryu}}, \bibinfo {author} {\bibfnamefont {E.}~\bibnamefont {Samson}},\ and\
  \bibinfo {author} {\bibfnamefont {M.~G.}\ \bibnamefont {Boshier}},\ }\href
  {https://www.nature.com/articles/s41467-020-17185-6} {\bibfield  {journal}
  {\bibinfo  {journal} {Nat. Commun.}\ }\textbf {\bibinfo {volume} {11}},\
  \bibinfo {pages} {3338} (\bibinfo {year} {2020})}\BibitemShut {NoStop}%
\bibitem [{\citenamefont {Amico}\ \emph {et~al.}(2021)\citenamefont {Amico},
  \citenamefont {Boshier}, \citenamefont {Birkl}, \citenamefont {Minguzzi},
  \citenamefont {Miniatura}, \citenamefont {Kwek}, \citenamefont {Aghamalyan},
  \citenamefont {Ahufinger}, \citenamefont {Anderson}, \citenamefont {Andrei}
  \emph {et~al.}}]{Amico2021}%
  \BibitemOpen
  \bibfield  {author} {\bibinfo {author} {\bibfnamefont {L.}~\bibnamefont
  {Amico}}, \bibinfo {author} {\bibfnamefont {M.}~\bibnamefont {Boshier}},
  \bibinfo {author} {\bibfnamefont {G.}~\bibnamefont {Birkl}}, \bibinfo
  {author} {\bibfnamefont {A.}~\bibnamefont {Minguzzi}}, \bibinfo {author}
  {\bibfnamefont {C.}~\bibnamefont {Miniatura}}, \bibinfo {author}
  {\bibfnamefont {L.-C.}\ \bibnamefont {Kwek}}, \bibinfo {author}
  {\bibfnamefont {D.}~\bibnamefont {Aghamalyan}}, \bibinfo {author}
  {\bibfnamefont {V.}~\bibnamefont {Ahufinger}}, \bibinfo {author}
  {\bibfnamefont {D.}~\bibnamefont {Anderson}}, \bibinfo {author}
  {\bibfnamefont {N.}~\bibnamefont {Andrei}}, \emph {et~al.},\ }\href
  {https://avs.scitation.org/doi/10.1116/5.0026178} {\bibfield  {journal}
  {\bibinfo  {journal} {AVS Quantum Science}\ }\textbf {\bibinfo {volume}
  {3}},\ \bibinfo {pages} {039201} (\bibinfo {year} {2021})}\BibitemShut
  {NoStop}%
\bibitem [{\citenamefont {Pandey}\ \emph {et~al.}(2021)\citenamefont {Pandey},
  \citenamefont {Mas}, \citenamefont {Vasilakis},\ and\ \citenamefont {von
  Klitzing}}]{Pandey2021}%
  \BibitemOpen
  \bibfield  {author} {\bibinfo {author} {\bibfnamefont {S.}~\bibnamefont
  {Pandey}}, \bibinfo {author} {\bibfnamefont {H.}~\bibnamefont {Mas}},
  \bibinfo {author} {\bibfnamefont {G.}~\bibnamefont {Vasilakis}},\ and\
  \bibinfo {author} {\bibfnamefont {W.}~\bibnamefont {von Klitzing}},\ }\href
  {https://journals.aps.org/prl/abstract/10.1103/PhysRevLett.126.170402}
  {\bibfield  {journal} {\bibinfo  {journal} {Phys. Rev. Lett.}\ }\textbf
  {\bibinfo {volume} {126}},\ \bibinfo {pages} {170402} (\bibinfo {year}
  {2021})}\BibitemShut {NoStop}%
\bibitem [{\citenamefont {P{\'e}rez-Obiol}\ \emph {et~al.}(2022)\citenamefont
  {P{\'e}rez-Obiol}, \citenamefont {Polo},\ and\ \citenamefont
  {Amico}}]{Perez2022}%
  \BibitemOpen
  \bibfield  {author} {\bibinfo {author} {\bibfnamefont {A.}~\bibnamefont
  {P{\'e}rez-Obiol}}, \bibinfo {author} {\bibfnamefont {J.}~\bibnamefont
  {Polo}},\ and\ \bibinfo {author} {\bibfnamefont {L.}~\bibnamefont {Amico}},\
  }\href
  {https://journals.aps.org/prresearch/abstract/10.1103/PhysRevResearch.4.L022038}
  {\bibfield  {journal} {\bibinfo  {journal} {Phys. Rev. Research}\ }\textbf
  {\bibinfo {volume} {4}},\ \bibinfo {pages} {L022038} (\bibinfo {year}
  {2022})}\BibitemShut {NoStop}%
\bibitem [{\citenamefont {Gr{\"u}n}\ \emph {et~al.}(2022)\citenamefont
  {Gr{\"u}n}, \citenamefont {Ymai}, \citenamefont {W}, \citenamefont {Tonel},
  \citenamefont {Foerster},\ and\ \citenamefont {Links}}]{Grun2022}%
  \BibitemOpen
  \bibfield  {author} {\bibinfo {author} {\bibfnamefont {D.}~\bibnamefont
  {Gr{\"u}n}}, \bibinfo {author} {\bibfnamefont {L.}~\bibnamefont {Ymai}},
  \bibinfo {author} {\bibfnamefont {K.~W.}\ \bibnamefont {W}}, \bibinfo
  {author} {\bibfnamefont {A.}~\bibnamefont {Tonel}}, \bibinfo {author}
  {\bibfnamefont {A.}~\bibnamefont {Foerster}},\ and\ \bibinfo {author}
  {\bibfnamefont {J.}~\bibnamefont {Links}},\ }\href
  {https://journals.aps.org/prl/abstract/10.1103/PhysRevLett.129.020401}
  {\bibfield  {journal} {\bibinfo  {journal} {Phys. Rev, Lett.}\ }\textbf
  {\bibinfo {volume} {129}},\ \bibinfo {pages} {020401} (\bibinfo {year}
  {2022})}\BibitemShut {NoStop}%
\bibitem [{\citenamefont {Kinoshita}\ \emph {et~al.}(2006)\citenamefont
  {Kinoshita}, \citenamefont {Wenger},\ and\ \citenamefont
  {Weiss}}]{Kinoshita2006}%
  \BibitemOpen
  \bibfield  {author} {\bibinfo {author} {\bibfnamefont {T.}~\bibnamefont
  {Kinoshita}}, \bibinfo {author} {\bibfnamefont {T.}~\bibnamefont {Wenger}},\
  and\ \bibinfo {author} {\bibfnamefont {D.~S.}\ \bibnamefont {Weiss}},\ }\href
  {https://www.nature.com/articles/nature04693} {\bibfield  {journal} {\bibinfo
   {journal} {Nature}\ }\textbf {\bibinfo {volume} {440}},\ \bibinfo {pages}
  {900} (\bibinfo {year} {2006})}\BibitemShut {NoStop}%
\bibitem [{\citenamefont {Hung}\ \emph {et~al.}(2013)\citenamefont {Hung},
  \citenamefont {Gurarie},\ and\ \citenamefont {Chin}}]{Hung2013}%
  \BibitemOpen
  \bibfield  {author} {\bibinfo {author} {\bibfnamefont {C.-L.}\ \bibnamefont
  {Hung}}, \bibinfo {author} {\bibfnamefont {V.}~\bibnamefont {Gurarie}},\ and\
  \bibinfo {author} {\bibfnamefont {C.}~\bibnamefont {Chin}},\ }\href
  {https://www.science.org/doi/full/10.1126/science.1237557?casa_token=abg8RX0WO-MAAAAA%3AND1GBLE4C-MEmBQTtT0_a7c0QQYv7-t-843XjcdES0kb0Nku4-0pQJ65k8hhtphTtoLfsvszAuxlTjo}
  {\bibfield  {journal} {\bibinfo  {journal} {Science}\ }\textbf {\bibinfo
  {volume} {341}},\ \bibinfo {pages} {1213} (\bibinfo {year}
  {2013})}\BibitemShut {NoStop}%
\bibitem [{\citenamefont {Pigneur}\ \emph {et~al.}(2018)\citenamefont
  {Pigneur}, \citenamefont {Berrada}, \citenamefont {Bonneau}, \citenamefont
  {Schumm}, \citenamefont {Demler},\ and\ \citenamefont
  {Schmiedmayer}}]{Pigneur2018}%
  \BibitemOpen
  \bibfield  {author} {\bibinfo {author} {\bibfnamefont {M.}~\bibnamefont
  {Pigneur}}, \bibinfo {author} {\bibfnamefont {T.}~\bibnamefont {Berrada}},
  \bibinfo {author} {\bibfnamefont {M.}~\bibnamefont {Bonneau}}, \bibinfo
  {author} {\bibfnamefont {T.}~\bibnamefont {Schumm}}, \bibinfo {author}
  {\bibfnamefont {E.}~\bibnamefont {Demler}},\ and\ \bibinfo {author}
  {\bibfnamefont {J.}~\bibnamefont {Schmiedmayer}},\ }\href
  {https://link.aps.org/doi/10.1103/PhysRevLett.120.173601} {\bibfield
  {journal} {\bibinfo  {journal} {Phys. Rev. Lett.}\ }\textbf {\bibinfo
  {volume} {120}},\ \bibinfo {pages} {173601} (\bibinfo {year}
  {2018})}\BibitemShut {NoStop}%
\bibitem [{\citenamefont {Chen}\ \emph {et~al.}(2019)\citenamefont {Chen},
  \citenamefont {Tang}, \citenamefont {Austin}, \citenamefont {Shaw},
  \citenamefont {Zhao},\ and\ \citenamefont {Liu}}]{Chen2019}%
  \BibitemOpen
  \bibfield  {author} {\bibinfo {author} {\bibfnamefont {Z.}~\bibnamefont
  {Chen}}, \bibinfo {author} {\bibfnamefont {T.}~\bibnamefont {Tang}}, \bibinfo
  {author} {\bibfnamefont {J.}~\bibnamefont {Austin}}, \bibinfo {author}
  {\bibfnamefont {Z.}~\bibnamefont {Shaw}}, \bibinfo {author} {\bibfnamefont
  {L.}~\bibnamefont {Zhao}},\ and\ \bibinfo {author} {\bibfnamefont
  {Y.}~\bibnamefont {Liu}},\ }\href
  {https://journals.aps.org/prl/abstract/10.1103/PhysRevLett.123.113002}
  {\bibfield  {journal} {\bibinfo  {journal} {Phys. Rev. Lett}\ }\textbf
  {\bibinfo {volume} {123}},\ \bibinfo {pages} {113002} (\bibinfo {year}
  {2019})}\BibitemShut {NoStop}%
\bibitem [{\citenamefont {Lamporesi}\ \emph {et~al.}(2013)\citenamefont
  {Lamporesi}, \citenamefont {Donadello}, \citenamefont {Serafini},
  \citenamefont {Dalfovo},\ and\ \citenamefont {Ferrari}}]{Lamporesi2013}%
  \BibitemOpen
  \bibfield  {author} {\bibinfo {author} {\bibfnamefont {G.}~\bibnamefont
  {Lamporesi}}, \bibinfo {author} {\bibfnamefont {S.}~\bibnamefont
  {Donadello}}, \bibinfo {author} {\bibfnamefont {S.}~\bibnamefont {Serafini}},
  \bibinfo {author} {\bibfnamefont {F.}~\bibnamefont {Dalfovo}},\ and\ \bibinfo
  {author} {\bibfnamefont {G.}~\bibnamefont {Ferrari}},\ }\href
  {https://www.nature.com/articles/nphys2734} {\bibfield  {journal} {\bibinfo
  {journal} {Nat. Phys.}\ }\textbf {\bibinfo {volume} {9}},\ \bibinfo {pages}
  {656} (\bibinfo {year} {2013})}\BibitemShut {NoStop}%
\bibitem [{\citenamefont {Krinner}\ \emph {et~al.}(2015)\citenamefont
  {Krinner}, \citenamefont {Stadler}, \citenamefont {Husmann}, \citenamefont
  {Brantut},\ and\ \citenamefont {Esslinger}}]{Krinner2015}%
  \BibitemOpen
  \bibfield  {author} {\bibinfo {author} {\bibfnamefont {S.}~\bibnamefont
  {Krinner}}, \bibinfo {author} {\bibfnamefont {D.}~\bibnamefont {Stadler}},
  \bibinfo {author} {\bibfnamefont {D.}~\bibnamefont {Husmann}}, \bibinfo
  {author} {\bibfnamefont {J.-P.}\ \bibnamefont {Brantut}},\ and\ \bibinfo
  {author} {\bibfnamefont {T.}~\bibnamefont {Esslinger}},\ }\href
  {https://www.nature.com/articles/nature14049} {\bibfield  {journal} {\bibinfo
   {journal} {Nature}\ }\textbf {\bibinfo {volume} {517}},\ \bibinfo {pages}
  {64} (\bibinfo {year} {2015})}\BibitemShut {NoStop}%
\bibitem [{\citenamefont {Erne}\ \emph {et~al.}(2018)\citenamefont {Erne},
  \citenamefont {B{\"u}cker}, \citenamefont {Gasenzer}, \citenamefont
  {Berges},\ and\ \citenamefont {Schmiedmayer}}]{Erne2018}%
  \BibitemOpen
  \bibfield  {author} {\bibinfo {author} {\bibfnamefont {S.}~\bibnamefont
  {Erne}}, \bibinfo {author} {\bibfnamefont {R.}~\bibnamefont {B{\"u}cker}},
  \bibinfo {author} {\bibfnamefont {T.}~\bibnamefont {Gasenzer}}, \bibinfo
  {author} {\bibfnamefont {J.}~\bibnamefont {Berges}},\ and\ \bibinfo {author}
  {\bibfnamefont {J.}~\bibnamefont {Schmiedmayer}},\ }\href
  {https://www.nature.com/articles/s41586-018-0667-0} {\bibfield  {journal}
  {\bibinfo  {journal} {Nature}\ }\textbf {\bibinfo {volume} {563}},\ \bibinfo
  {pages} {225} (\bibinfo {year} {2018})}\BibitemShut {NoStop}%
\bibitem [{\citenamefont {Eigen}\ \emph {et~al.}(2018)\citenamefont {Eigen},
  \citenamefont {Glidden}, \citenamefont {Lopes}, \citenamefont {Cornell},
  \citenamefont {Smith},\ and\ \citenamefont {Hadzibabic}}]{Eigen2018}%
  \BibitemOpen
  \bibfield  {author} {\bibinfo {author} {\bibfnamefont {C.}~\bibnamefont
  {Eigen}}, \bibinfo {author} {\bibfnamefont {J.~A.}\ \bibnamefont {Glidden}},
  \bibinfo {author} {\bibfnamefont {R.}~\bibnamefont {Lopes}}, \bibinfo
  {author} {\bibfnamefont {E.~A.}\ \bibnamefont {Cornell}}, \bibinfo {author}
  {\bibfnamefont {R.~P.}\ \bibnamefont {Smith}},\ and\ \bibinfo {author}
  {\bibfnamefont {Z.}~\bibnamefont {Hadzibabic}},\ }\href
  {https://www.nature.com/articles/s41586-018-0674-1} {\bibfield  {journal}
  {\bibinfo  {journal} {Nature}\ }\textbf {\bibinfo {volume} {563}},\ \bibinfo
  {pages} {221} (\bibinfo {year} {2018})}\BibitemShut {NoStop}%
\bibitem [{\citenamefont {Navon}\ \emph {et~al.}(2016)\citenamefont {Navon},
  \citenamefont {Gaunt}, \citenamefont {Smith},\ and\ \citenamefont
  {Hadzibabic}}]{navon2016emergence}%
  \BibitemOpen
  \bibfield  {author} {\bibinfo {author} {\bibfnamefont {N.}~\bibnamefont
  {Navon}}, \bibinfo {author} {\bibfnamefont {A.~L.}\ \bibnamefont {Gaunt}},
  \bibinfo {author} {\bibfnamefont {R.~P.}\ \bibnamefont {Smith}},\ and\
  \bibinfo {author} {\bibfnamefont {Z.}~\bibnamefont {Hadzibabic}},\ }\href
  {https://www.nature.com/articles/nature20114} {\bibfield  {journal} {\bibinfo
   {journal} {Nature}\ }\textbf {\bibinfo {volume} {539}},\ \bibinfo {pages}
  {72} (\bibinfo {year} {2016})}\BibitemShut {NoStop}%
\bibitem [{\citenamefont {Navon}\ \emph {et~al.}(2019)\citenamefont {Navon},
  \citenamefont {Eigen}, \citenamefont {Zhang}, \citenamefont {Lopes},
  \citenamefont {Gaunt}, \citenamefont {Fujimoto}, \citenamefont {Tsubota},
  \citenamefont {Smith},\ and\ \citenamefont
  {Hadzibabic}}]{navon2019synthetic}%
  \BibitemOpen
  \bibfield  {author} {\bibinfo {author} {\bibfnamefont {N.}~\bibnamefont
  {Navon}}, \bibinfo {author} {\bibfnamefont {C.}~\bibnamefont {Eigen}},
  \bibinfo {author} {\bibfnamefont {J.}~\bibnamefont {Zhang}}, \bibinfo
  {author} {\bibfnamefont {R.}~\bibnamefont {Lopes}}, \bibinfo {author}
  {\bibfnamefont {A.~L.}\ \bibnamefont {Gaunt}}, \bibinfo {author}
  {\bibfnamefont {K.}~\bibnamefont {Fujimoto}}, \bibinfo {author}
  {\bibfnamefont {M.}~\bibnamefont {Tsubota}}, \bibinfo {author} {\bibfnamefont
  {R.~P.}\ \bibnamefont {Smith}},\ and\ \bibinfo {author} {\bibfnamefont
  {Z.}~\bibnamefont {Hadzibabic}},\ }\href
  {https://www.science.org/doi/full/10.1126/science.aau6103?casa_token=PQ_3eNfrmCAAAAAA:vy1ux3phekgKl1a5GSBGuA23mW8VzkQTcyHcw9v7CxCWQCq9Nk3X_FGSLtFyvNNNOajZ7CCjR_9dphM}
  {\bibfield  {journal} {\bibinfo  {journal} {Science}\ }\textbf {\bibinfo
  {volume} {366}},\ \bibinfo {pages} {382} (\bibinfo {year}
  {2019})}\BibitemShut {NoStop}%
\bibitem [{\citenamefont {Reeves}\ \emph {et~al.}(2022)\citenamefont {Reeves},
  \citenamefont {Goddard-Lee}, \citenamefont {Gauthier}, \citenamefont
  {Stockdale}, \citenamefont {Salman}, \citenamefont {Edmonds}, \citenamefont
  {Yu}, \citenamefont {Bradley}, \citenamefont {Baker}, \citenamefont
  {Rubinsztein-Dunlop} \emph {et~al.}}]{reeves2022turbulent}%
  \BibitemOpen
  \bibfield  {author} {\bibinfo {author} {\bibfnamefont {M.~T.}\ \bibnamefont
  {Reeves}}, \bibinfo {author} {\bibfnamefont {K.}~\bibnamefont {Goddard-Lee}},
  \bibinfo {author} {\bibfnamefont {G.}~\bibnamefont {Gauthier}}, \bibinfo
  {author} {\bibfnamefont {O.~R.}\ \bibnamefont {Stockdale}}, \bibinfo {author}
  {\bibfnamefont {H.}~\bibnamefont {Salman}}, \bibinfo {author} {\bibfnamefont
  {T.}~\bibnamefont {Edmonds}}, \bibinfo {author} {\bibfnamefont
  {X.}~\bibnamefont {Yu}}, \bibinfo {author} {\bibfnamefont {A.~S.}\
  \bibnamefont {Bradley}}, \bibinfo {author} {\bibfnamefont {M.}~\bibnamefont
  {Baker}}, \bibinfo {author} {\bibfnamefont {H.}~\bibnamefont
  {Rubinsztein-Dunlop}}, \emph {et~al.},\ }\href
  {https://journals.aps.org/prx/abstract/10.1103/PhysRevX.12.011031} {\bibfield
   {journal} {\bibinfo  {journal} {Phys. Rev. X}\ }\textbf {\bibinfo {volume}
  {12}},\ \bibinfo {pages} {011031} (\bibinfo {year} {2022})}\BibitemShut
  {NoStop}%
\bibitem [{\citenamefont {Labouvie}\ \emph {et~al.}(2015)\citenamefont
  {Labouvie}, \citenamefont {Santra}, \citenamefont {Heun}, \citenamefont
  {Wimberger},\ and\ \citenamefont {Ott}}]{Labouvie2015}%
  \BibitemOpen
  \bibfield  {author} {\bibinfo {author} {\bibfnamefont {R.}~\bibnamefont
  {Labouvie}}, \bibinfo {author} {\bibfnamefont {B.}~\bibnamefont {Santra}},
  \bibinfo {author} {\bibfnamefont {S.}~\bibnamefont {Heun}}, \bibinfo {author}
  {\bibfnamefont {S.}~\bibnamefont {Wimberger}},\ and\ \bibinfo {author}
  {\bibfnamefont {H.}~\bibnamefont {Ott}},\ }\href
  {https://journals.aps.org/prl/abstract/10.1103/PhysRevLett.115.050601}
  {\bibfield  {journal} {\bibinfo  {journal} {Phys. Rev. Lett.}\ }\textbf
  {\bibinfo {volume} {115}},\ \bibinfo {pages} {050601} (\bibinfo {year}
  {2015})}\BibitemShut {NoStop}%
\bibitem [{\citenamefont {Labouvie}\ \emph {et~al.}(2016)\citenamefont
  {Labouvie}, \citenamefont {Santra}, \citenamefont {Heun},\ and\ \citenamefont
  {Ott}}]{Labouvie2016}%
  \BibitemOpen
  \bibfield  {author} {\bibinfo {author} {\bibfnamefont {R.}~\bibnamefont
  {Labouvie}}, \bibinfo {author} {\bibfnamefont {B.}~\bibnamefont {Santra}},
  \bibinfo {author} {\bibfnamefont {S.}~\bibnamefont {Heun}},\ and\ \bibinfo
  {author} {\bibfnamefont {H.}~\bibnamefont {Ott}},\ }\href
  {https://journals.aps.org/prl/abstract/10.1103/PhysRevLett.116.235302}
  {\bibfield  {journal} {\bibinfo  {journal} {Phys. Rev. Lett.}\ }\textbf
  {\bibinfo {volume} {116}},\ \bibinfo {pages} {235302} (\bibinfo {year}
  {2016})}\BibitemShut {NoStop}%
\bibitem [{\citenamefont {Ceulemans}\ and\ \citenamefont
  {Wouters}(2023)}]{Ceulemans2023}%
  \BibitemOpen
  \bibfield  {author} {\bibinfo {author} {\bibfnamefont {R.}~\bibnamefont
  {Ceulemans}}\ and\ \bibinfo {author} {\bibfnamefont {M.}~\bibnamefont
  {Wouters}},\ }\href
  {https://journals.aps.org/pra/abstract/10.1103/PhysRevA.108.013314}
  {\bibfield  {journal} {\bibinfo  {journal} {Phys. Rev. A}\ }\textbf {\bibinfo
  {volume} {108}},\ \bibinfo {pages} {013314} (\bibinfo {year}
  {2023})}\BibitemShut {NoStop}%
\bibitem [{\citenamefont {Shaw}\ \emph
  {et~al.}(1992{\natexlab{a}})\citenamefont {Shaw}, \citenamefont {Mitin},
  \citenamefont {Sch{\"o}ll},\ and\ \citenamefont {Grubin}}]{Shaw1992_1}%
  \BibitemOpen
  \bibfield  {author} {\bibinfo {author} {\bibfnamefont {M.~P.}\ \bibnamefont
  {Shaw}}, \bibinfo {author} {\bibfnamefont {V.~V.}\ \bibnamefont {Mitin}},
  \bibinfo {author} {\bibfnamefont {E.}~\bibnamefont {Sch{\"o}ll}},\ and\
  \bibinfo {author} {\bibfnamefont {H.~L.}\ \bibnamefont {Grubin}},\ }\bibinfo
  {title} {Introduction},\ in\ \href
  {https://doi.org/10.1007/978-1-4899-2344-8_1} {\emph {\bibinfo {booktitle}
  {The Physics of Instabilities in Solid State Electron Devices}}},\ \bibinfo
  {editor} {edited by\ \bibinfo {editor} {\bibfnamefont {M.~P.}\ \bibnamefont
  {Shaw}}, \bibinfo {editor} {\bibfnamefont {V.~V.}\ \bibnamefont {Mitin}},
  \bibinfo {editor} {\bibfnamefont {E.}~\bibnamefont {Sch{\"o}ll}},\ and\
  \bibinfo {editor} {\bibfnamefont {H.~L.}\ \bibnamefont {Grubin}}}\ (\bibinfo
  {publisher} {Springer US},\ \bibinfo {address} {Boston, MA},\ \bibinfo {year}
  {1992})\ pp.\ \bibinfo {pages} {1--70}\BibitemShut {NoStop}%
\bibitem [{\citenamefont {Perrin}\ \emph {et~al.}(2014)\citenamefont {Perrin},
  \citenamefont {Frisenda}, \citenamefont {Koole}, \citenamefont {Seldenthuis},
  \citenamefont {Gil}, \citenamefont {Valkenier}, \citenamefont {Hummelen},
  \citenamefont {Renaud}, \citenamefont {Grozema}, \citenamefont {Thijssen}
  \emph {et~al.}}]{Perrin2014}%
  \BibitemOpen
  \bibfield  {author} {\bibinfo {author} {\bibfnamefont {M.~L.}\ \bibnamefont
  {Perrin}}, \bibinfo {author} {\bibfnamefont {R.}~\bibnamefont {Frisenda}},
  \bibinfo {author} {\bibfnamefont {M.}~\bibnamefont {Koole}}, \bibinfo
  {author} {\bibfnamefont {J.~S.}\ \bibnamefont {Seldenthuis}}, \bibinfo
  {author} {\bibfnamefont {J.~A.~C.}\ \bibnamefont {Gil}}, \bibinfo {author}
  {\bibfnamefont {H.}~\bibnamefont {Valkenier}}, \bibinfo {author}
  {\bibfnamefont {J.~C.}\ \bibnamefont {Hummelen}}, \bibinfo {author}
  {\bibfnamefont {N.}~\bibnamefont {Renaud}}, \bibinfo {author} {\bibfnamefont
  {F.~C.}\ \bibnamefont {Grozema}}, \bibinfo {author} {\bibfnamefont {J.~M.}\
  \bibnamefont {Thijssen}}, \emph {et~al.},\ }\href
  {https://www.nature.com/articles/nnano.2014.177} {\bibfield  {journal}
  {\bibinfo  {journal} {Nat. Nanotechnol.}\ }\textbf {\bibinfo {volume} {9}},\
  \bibinfo {pages} {830} (\bibinfo {year} {2014})}\BibitemShut {NoStop}%
\bibitem [{\citenamefont {Britnell}\ \emph {et~al.}(2013)\citenamefont
  {Britnell}, \citenamefont {Gorbachev}, \citenamefont {Geim}, \citenamefont
  {Ponomarenko}, \citenamefont {Mishchenko}, \citenamefont {Greenaway},
  \citenamefont {Fromhold}, \citenamefont {Novoselov},\ and\ \citenamefont
  {Eaves}}]{Britnell2013}%
  \BibitemOpen
  \bibfield  {author} {\bibinfo {author} {\bibfnamefont {L.}~\bibnamefont
  {Britnell}}, \bibinfo {author} {\bibfnamefont {R.}~\bibnamefont {Gorbachev}},
  \bibinfo {author} {\bibfnamefont {A.}~\bibnamefont {Geim}}, \bibinfo {author}
  {\bibfnamefont {L.}~\bibnamefont {Ponomarenko}}, \bibinfo {author}
  {\bibfnamefont {A.}~\bibnamefont {Mishchenko}}, \bibinfo {author}
  {\bibfnamefont {M.}~\bibnamefont {Greenaway}}, \bibinfo {author}
  {\bibfnamefont {T.}~\bibnamefont {Fromhold}}, \bibinfo {author}
  {\bibfnamefont {K.}~\bibnamefont {Novoselov}},\ and\ \bibinfo {author}
  {\bibfnamefont {L.}~\bibnamefont {Eaves}},\ }\href
  {https://www.nature.com/articles/ncomms2817} {\bibfield  {journal} {\bibinfo
  {journal} {Nat. Commun.}\ }\textbf {\bibinfo {volume} {4}},\ \bibinfo {pages}
  {1794} (\bibinfo {year} {2013})}\BibitemShut {NoStop}%
\bibitem [{\citenamefont {Volkov}\ and\ \citenamefont
  {Kogan}(1969)}]{Volkov1969}%
  \BibitemOpen
  \bibfield  {author} {\bibinfo {author} {\bibfnamefont {A.~F.}\ \bibnamefont
  {Volkov}}\ and\ \bibinfo {author} {\bibfnamefont {S.~M.}\ \bibnamefont
  {Kogan}},\ }\href {https://doi.org/10.1070/pu1969v011n06abeh003780}
  {\bibfield  {journal} {\bibinfo  {journal} {Soviet Physics Uspekhi}\ }\textbf
  {\bibinfo {volume} {11}},\ \bibinfo {pages} {881} (\bibinfo {year}
  {1969})}\BibitemShut {NoStop}%
\bibitem [{\citenamefont {Pamplin}(1970)}]{Pamplin1970}%
  \BibitemOpen
  \bibfield  {author} {\bibinfo {author} {\bibfnamefont {B.~R.}\ \bibnamefont
  {Pamplin}},\ }\href
  {https://www.tandfonline.com/doi/abs/10.1080/00107517008204806?casa_token=zgeud40w4wAAAAAA:0b8GVpKw8taCwUQaTo8hOtUMNGZtk3luf2JJxRMgChQqQ0L4H-vhKZAzEDd1NVUMC1lEnFKkJx3JzQ}
  {\bibfield  {journal} {\bibinfo  {journal} {Contemp. Phys}\ }\textbf
  {\bibinfo {volume} {11}},\ \bibinfo {pages} {1} (\bibinfo {year}
  {1970})}\BibitemShut {NoStop}%
\bibitem [{\citenamefont {Brantut}\ \emph {et~al.}(2013)\citenamefont
  {Brantut}, \citenamefont {Grenier}, \citenamefont {Meineke}, \citenamefont
  {Stadler}, \citenamefont {Krinner}, \citenamefont {Kollath}, \citenamefont
  {Esslinger},\ and\ \citenamefont {Georges}}]{Brantut2013}%
  \BibitemOpen
  \bibfield  {author} {\bibinfo {author} {\bibfnamefont {J.-P.}\ \bibnamefont
  {Brantut}}, \bibinfo {author} {\bibfnamefont {C.}~\bibnamefont {Grenier}},
  \bibinfo {author} {\bibfnamefont {J.}~\bibnamefont {Meineke}}, \bibinfo
  {author} {\bibfnamefont {D.}~\bibnamefont {Stadler}}, \bibinfo {author}
  {\bibfnamefont {S.}~\bibnamefont {Krinner}}, \bibinfo {author} {\bibfnamefont
  {C.}~\bibnamefont {Kollath}}, \bibinfo {author} {\bibfnamefont
  {T.}~\bibnamefont {Esslinger}},\ and\ \bibinfo {author} {\bibfnamefont
  {A.}~\bibnamefont {Georges}},\ }\href
  {https://www.science.org/doi/full/10.1126/science.1242308?casa_token=k5eTpncR0MUAAAAA%3Ax5ikFNHCkCpUn2Wf-JTNEQ_rAOwNXEZV1yWkR8IuJQW0lB0pNDDTHutvBqmv4vCmST1F7BE-pQ86h6o}
  {\bibfield  {journal} {\bibinfo  {journal} {Science}\ }\textbf {\bibinfo
  {volume} {342}},\ \bibinfo {pages} {713} (\bibinfo {year}
  {2013})}\BibitemShut {NoStop}%
\bibitem [{\citenamefont {Chien}\ \emph {et~al.}(2015)\citenamefont {Chien},
  \citenamefont {Peotta},\ and\ \citenamefont {Di~Ventra}}]{Chien2015}%
  \BibitemOpen
  \bibfield  {author} {\bibinfo {author} {\bibfnamefont {C.-C.}\ \bibnamefont
  {Chien}}, \bibinfo {author} {\bibfnamefont {S.}~\bibnamefont {Peotta}},\ and\
  \bibinfo {author} {\bibfnamefont {M.}~\bibnamefont {Di~Ventra}},\ }\href
  {https://www.nature.com/articles/nphys3531} {\bibfield  {journal} {\bibinfo
  {journal} {Nat. Phys.}\ }\textbf {\bibinfo {volume} {11}},\ \bibinfo {pages}
  {998} (\bibinfo {year} {2015})}\BibitemShut {NoStop}%
\bibitem [{\citenamefont {Mink}\ \emph {et~al.}(2022)\citenamefont {Mink},
  \citenamefont {Pelster}, \citenamefont {Benary}, \citenamefont {Ott},\ and\
  \citenamefont {Fleischhauer}}]{Mink2022}%
  \BibitemOpen
  \bibfield  {author} {\bibinfo {author} {\bibfnamefont {C.}~\bibnamefont
  {Mink}}, \bibinfo {author} {\bibfnamefont {A.}~\bibnamefont {Pelster}},
  \bibinfo {author} {\bibfnamefont {J.}~\bibnamefont {Benary}}, \bibinfo
  {author} {\bibfnamefont {H.}~\bibnamefont {Ott}},\ and\ \bibinfo {author}
  {\bibfnamefont {M.}~\bibnamefont {Fleischhauer}},\ }\href
  {https://scipost.org/10.21468/SciPostPhys.12.2.051} {\bibfield  {journal}
  {\bibinfo  {journal} {SciPost Phys.}\ }\textbf {\bibinfo {volume} {12}},\
  \bibinfo {pages} {051} (\bibinfo {year} {2022})}\BibitemShut {NoStop}%
\bibitem [{\citenamefont {Olsen}\ and\ \citenamefont
  {Corney}(2016)}]{Olsen2016}%
  \BibitemOpen
  \bibfield  {author} {\bibinfo {author} {\bibfnamefont {M.~K.}\ \bibnamefont
  {Olsen}}\ and\ \bibinfo {author} {\bibfnamefont {J.~F.}\ \bibnamefont
  {Corney}},\ }\href
  {https://journals.aps.org/pra/abstract/10.1103/PhysRevA.94.033605} {\bibfield
   {journal} {\bibinfo  {journal} {Phys. Rev. A}\ }\textbf {\bibinfo {volume}
  {94}},\ \bibinfo {pages} {033605} (\bibinfo {year} {2016})}\BibitemShut
  {NoStop}%
\bibitem [{\citenamefont {Fischer}\ and\ \citenamefont
  {Wimberger}(2017)}]{Fischer2017}%
  \BibitemOpen
  \bibfield  {author} {\bibinfo {author} {\bibfnamefont {D.}~\bibnamefont
  {Fischer}}\ and\ \bibinfo {author} {\bibfnamefont {S.}~\bibnamefont
  {Wimberger}},\ }\href
  {https://onlinelibrary.wiley.com/doi/abs/10.1002/andp.201600327} {\bibfield
  {journal} {\bibinfo  {journal} {Ann. Phys.}\ }\textbf {\bibinfo {volume}
  {529}},\ \bibinfo {pages} {1600327} (\bibinfo {year} {2017})}\BibitemShut
  {NoStop}%
\bibitem [{\citenamefont {Steel}\ \emph {et~al.}(1998)\citenamefont {Steel},
  \citenamefont {Olsen}, \citenamefont {Plimak}, \citenamefont {Drummond},
  \citenamefont {Tan}, \citenamefont {Collett}, \citenamefont {Walls},\ and\
  \citenamefont {Graham}}]{Steel1998}%
  \BibitemOpen
  \bibfield  {author} {\bibinfo {author} {\bibfnamefont {M.~J.}\ \bibnamefont
  {Steel}}, \bibinfo {author} {\bibfnamefont {M.~K.}\ \bibnamefont {Olsen}},
  \bibinfo {author} {\bibfnamefont {L.~I.}\ \bibnamefont {Plimak}}, \bibinfo
  {author} {\bibfnamefont {P.~D.}\ \bibnamefont {Drummond}}, \bibinfo {author}
  {\bibfnamefont {S.~M.}\ \bibnamefont {Tan}}, \bibinfo {author} {\bibfnamefont
  {M.~J.}\ \bibnamefont {Collett}}, \bibinfo {author} {\bibfnamefont {D.~F.}\
  \bibnamefont {Walls}},\ and\ \bibinfo {author} {\bibfnamefont
  {R.}~\bibnamefont {Graham}},\ }\href
  {https://journals.aps.org/pra/abstract/10.1103/PhysRevA.58.4824} {\bibfield
  {journal} {\bibinfo  {journal} {Phys. Rev. A}\ }\textbf {\bibinfo {volume}
  {58}},\ \bibinfo {pages} {4824} (\bibinfo {year} {1998})}\BibitemShut
  {NoStop}%
\bibitem [{\citenamefont {Polkovnikov}(2003)}]{Polkovnikov2003a}%
  \BibitemOpen
  \bibfield  {author} {\bibinfo {author} {\bibfnamefont {A.}~\bibnamefont
  {Polkovnikov}},\ }\href
  {https://journals.aps.org/pra/abstract/10.1103/PhysRevA.68.053604} {\bibfield
   {journal} {\bibinfo  {journal} {Phys. Rev. A}\ }\textbf {\bibinfo {volume}
  {68}},\ \bibinfo {pages} {053604} (\bibinfo {year} {2003})}\BibitemShut
  {NoStop}%
\bibitem [{\citenamefont {Blakie}\ \emph {et~al.}(2008)\citenamefont {Blakie},
  \citenamefont {Bradley}, \citenamefont {Davis}, \citenamefont {Ballagh},\
  and\ \citenamefont {Gardiner}}]{Blakie2008}%
  \BibitemOpen
  \bibfield  {author} {\bibinfo {author} {\bibfnamefont {P.~B.}\ \bibnamefont
  {Blakie}}, \bibinfo {author} {\bibfnamefont {A.~S.}\ \bibnamefont {Bradley}},
  \bibinfo {author} {\bibfnamefont {M.~J.}\ \bibnamefont {Davis}}, \bibinfo
  {author} {\bibfnamefont {R.~J.}\ \bibnamefont {Ballagh}},\ and\ \bibinfo
  {author} {\bibfnamefont {C.~W.}\ \bibnamefont {Gardiner}},\ }\href
  {https://www.tandfonline.com/doi/full/10.1080/00018730802564254?casa_token=bHbRxWesdhEAAAAA%3AcAKKqK6F1YQQdqqnCXZlKqtMITGU65aOGRg0HgBVnmtCadskeNNkBOK5L7v9YMyh0Z50sAeFTu-Rqg}
  {\bibfield  {journal} {\bibinfo  {journal} {Adv. Phys}\ }\textbf {\bibinfo
  {volume} {57}},\ \bibinfo {pages} {363} (\bibinfo {year} {2008})}\BibitemShut
  {NoStop}%
\bibitem [{SM()}]{SM}%
  \BibitemOpen
  \href@noop {} {}\bibinfo {note} {See Supplemental Material for full details
  of the simulation parameters, a discussion and demonstration of the
  robustness of the solutions to the initial conditions, an analysis of the
  condensation dynamics of the central site, and a numerical experiment
  demonstrating that the system is far-from-equilibrium. It also includes
  Refs.~\cite{LabouvieThesis,Boyd2001,Arzamasovs2017,Zwerger2003,Sinatra2008}.}\BibitemShut
  {Stop}%
\bibitem [{\citenamefont {Dennis}\ \emph {et~al.}(2013)\citenamefont {Dennis},
  \citenamefont {Hope},\ and\ \citenamefont {Johnsson}}]{xmds2}%
  \BibitemOpen
  \bibfield  {author} {\bibinfo {author} {\bibfnamefont {G.~R.}\ \bibnamefont
  {Dennis}}, \bibinfo {author} {\bibfnamefont {J.~J.}\ \bibnamefont {Hope}},\
  and\ \bibinfo {author} {\bibfnamefont {M.~T.}\ \bibnamefont {Johnsson}},\
  }\href {https://www.sciencedirect.com/science/article/pii/S0010465512002822}
  {\bibfield  {journal} {\bibinfo  {journal} {Comput. Phys. Commun}\ }\textbf
  {\bibinfo {volume} {184}},\ \bibinfo {pages} {201} (\bibinfo {year}
  {2013})}\BibitemShut {NoStop}%
\bibitem [{nor()}]{normalisation_of_trajectories}%
  \BibitemOpen
  \href@noop {} {}\bibinfo {note} {In order to sensibly compare the atom number
  for individual trajectories of the Truncated Wigner Approximation to the mean
  atom number, it is necessary to subtract half an atom per mode from the
  normalisation of the trajectory in order to account for the initial quantum
  noise.}\BibitemShut {Stop}%
\bibitem [{\citenamefont {Pereverzev}\ \emph {et~al.}(1997)\citenamefont
  {Pereverzev}, \citenamefont {Loshak}, \citenamefont {Backhaus}, \citenamefont
  {Davis},\ and\ \citenamefont {Packard}}]{Pereverzev1997}%
  \BibitemOpen
  \bibfield  {author} {\bibinfo {author} {\bibfnamefont {S.~V.}\ \bibnamefont
  {Pereverzev}}, \bibinfo {author} {\bibfnamefont {A.}~\bibnamefont {Loshak}},
  \bibinfo {author} {\bibfnamefont {S.}~\bibnamefont {Backhaus}}, \bibinfo
  {author} {\bibfnamefont {J.~C.}\ \bibnamefont {Davis}},\ and\ \bibinfo
  {author} {\bibfnamefont {R.~E.}\ \bibnamefont {Packard}},\ }\href
  {https://www.nature.com/articles/41277} {\bibfield  {journal} {\bibinfo
  {journal} {Nature}\ }\textbf {\bibinfo {volume} {388}},\ \bibinfo {pages}
  {449} (\bibinfo {year} {1997})}\BibitemShut {NoStop}%
\bibitem [{\citenamefont {Levy}\ \emph {et~al.}(2007)\citenamefont {Levy},
  \citenamefont {Lahoud}, \citenamefont {Shomroni},\ and\ \citenamefont
  {Steinhauer}}]{Levy2007}%
  \BibitemOpen
  \bibfield  {author} {\bibinfo {author} {\bibfnamefont {S.}~\bibnamefont
  {Levy}}, \bibinfo {author} {\bibfnamefont {E.}~\bibnamefont {Lahoud}},
  \bibinfo {author} {\bibfnamefont {I.}~\bibnamefont {Shomroni}},\ and\
  \bibinfo {author} {\bibfnamefont {J.}~\bibnamefont {Steinhauer}},\ }\href
  {https://www.nature.com/articles/nature06186} {\bibfield  {journal} {\bibinfo
   {journal} {Nature}\ }\textbf {\bibinfo {volume} {449}},\ \bibinfo {pages}
  {579} (\bibinfo {year} {2007})}\BibitemShut {NoStop}%
\bibitem [{Fig()}]{Fig3footnote}%
  \BibitemOpen
  \href@noop {} {}\bibinfo {note} {To compensate for the disproportionate
  effect of small sampling errors in the region $N_0 \lesssim N_f$, for $I_m$
  we perform a rolling average over a time of $\Delta t= 10\omega_r^{-1}$. In
  (b), the rolling average is conducted over a window of
  $25\omega_r^{-1}$.}\BibitemShut {Stop}%
\bibitem [{\citenamefont {Shaw}\ \emph
  {et~al.}(1992{\natexlab{b}})\citenamefont {Shaw}, \citenamefont {Mitin},
  \citenamefont {Sch{\"o}ll},\ and\ \citenamefont {Grubin}}]{Shaw1992_6}%
  \BibitemOpen
  \bibfield  {author} {\bibinfo {author} {\bibfnamefont {M.~P.}\ \bibnamefont
  {Shaw}}, \bibinfo {author} {\bibfnamefont {V.~V.}\ \bibnamefont {Mitin}},
  \bibinfo {author} {\bibfnamefont {E.}~\bibnamefont {Sch{\"o}ll}},\ and\
  \bibinfo {author} {\bibfnamefont {H.~L.}\ \bibnamefont {Grubin}},\ }\bibinfo
  {title} {Superconducting junctions},\ in\ \href
  {https://doi.org/10.1007/978-1-4899-2344-8_6} {\emph {\bibinfo {booktitle}
  {The Physics of Instabilities in Solid State Electron Devices}}},\ \bibinfo
  {editor} {edited by\ \bibinfo {editor} {\bibfnamefont {M.~P.}\ \bibnamefont
  {Shaw}}, \bibinfo {editor} {\bibfnamefont {V.~V.}\ \bibnamefont {Mitin}},
  \bibinfo {editor} {\bibfnamefont {E.}~\bibnamefont {Sch{\"o}ll}},\ and\
  \bibinfo {editor} {\bibfnamefont {H.~L.}\ \bibnamefont {Grubin}}}\ (\bibinfo
  {publisher} {Springer US},\ \bibinfo {address} {Boston, MA},\ \bibinfo {year}
  {1992})\ pp.\ \bibinfo {pages} {325--358}\BibitemShut {NoStop}%
\bibitem [{\citenamefont {Reeves}\ and\ \citenamefont
  {Davis}(2023)}]{Reeves2021}%
  \BibitemOpen
  \bibfield  {author} {\bibinfo {author} {\bibfnamefont {M.~T.}\ \bibnamefont
  {Reeves}}\ and\ \bibinfo {author} {\bibfnamefont {M.~J.}\ \bibnamefont
  {Davis}},\ }\href
  {https://www.scipost.org/SciPostPhys.15.2.068?acad_field_slug=all} {\bibfield
   {journal} {\bibinfo  {journal} {SciPost Phys.}\ }\textbf {\bibinfo {volume}
  {15}},\ \bibinfo {pages} {068} (\bibinfo {year} {2023})}\BibitemShut
  {NoStop}%
\bibitem [{\citenamefont {Labouvie}(2015)}]{LabouvieThesis}%
  \BibitemOpen
  \bibfield  {author} {\bibinfo {author} {\bibfnamefont {R.}~\bibnamefont
  {Labouvie}},\ }\href@noop {} {\bibfield  {journal} {\bibinfo  {journal} {PhD
  Thesis, \textit{Non-equilibrium dynamics in ultracold quantum gases with
  localized dissipation}, Technischen Universit\"{a}t Kaiserslautern}\ }
  (\bibinfo {year} {2015})}\BibitemShut {NoStop}%
\bibitem [{\citenamefont {Boyd}(2001)}]{Boyd2001}%
  \BibitemOpen
  \bibfield  {author} {\bibinfo {author} {\bibfnamefont {J.~P.}\ \bibnamefont
  {Boyd}},\ }\href {https://link.springer.com/book/9783540514879} {\emph
  {\bibinfo {title} {Chebyshev and Fourier spectral methods}}}\ (\bibinfo
  {publisher} {Dover Publications},\ \bibinfo {year} {2001})\BibitemShut
  {NoStop}%
\bibitem [{\citenamefont {Arzamasovs}\ and\ \citenamefont
  {Liu}(2017)}]{Arzamasovs2017}%
  \BibitemOpen
  \bibfield  {author} {\bibinfo {author} {\bibfnamefont {M.}~\bibnamefont
  {Arzamasovs}}\ and\ \bibinfo {author} {\bibfnamefont {B.}~\bibnamefont
  {Liu}},\ }\href
  {https://iopscience.iop.org/article/10.1088/1361-6404/aa8d2c/meta?casa_token=j7a4GYhfYj4AAAAA:xkEkN8d_4jSBuiKHykPrs23DpgCZGGil3o4YyN_WfAU9eWdKxYpWwqh52kDgbZq199UXX5gSbMg}
  {\bibfield  {journal} {\bibinfo  {journal} {Eur. J. Phys.}\ }\textbf
  {\bibinfo {volume} {38}},\ \bibinfo {pages} {065405} (\bibinfo {year}
  {2017})}\BibitemShut {NoStop}%
\bibitem [{\citenamefont {Zwerger}(2003)}]{Zwerger2003}%
  \BibitemOpen
  \bibfield  {author} {\bibinfo {author} {\bibfnamefont {W.}~\bibnamefont
  {Zwerger}},\ }\href
  {https://iopscience.iop.org/article/10.1088/1464-4266/5/2/352/meta?casa_token=B-7T-aTveLgAAAAA:XROBCEGbM9d_0gvnhACZc_3BD1XEex6FwpalNWejAqe38ub8vbU2mdCuE7XHpP45nTmaVqG_e8I}
  {\bibfield  {journal} {\bibinfo  {journal} {J. Opt. B}\ }\textbf {\bibinfo
  {volume} {5}},\ \bibinfo {pages} {S9} (\bibinfo {year} {2003})}\BibitemShut
  {NoStop}%
\bibitem [{\citenamefont {Sinatra}\ and\ \citenamefont
  {Castin}(2008)}]{Sinatra2008}%
  \BibitemOpen
  \bibfield  {author} {\bibinfo {author} {\bibfnamefont {A.}~\bibnamefont
  {Sinatra}}\ and\ \bibinfo {author} {\bibfnamefont {Y.}~\bibnamefont
  {Castin}},\ }\href {https://doi.org/10.1103/PhysRevA.78.053615} {\bibfield
  {journal} {\bibinfo  {journal} {Phys. Rev. A}\ }\textbf {\bibinfo {volume}
  {78}},\ \bibinfo {pages} {053615} (\bibinfo {year} {2008})}\BibitemShut
  {NoStop}%
\end{thebibliography}

\end{document}

\title{Supplemental Material for ``Nonequilibrium  Transport in a Superfluid Josephson Junction Chain: Is There Negative Differential Conductivity?"}

\author{Samuel E. Begg}
\affiliation{Asia Pacific Center for Theoretical Physics, Pohang 37673, Korea}
\affiliation{Australian Research Council Centre of Excellence in Future Low-Energy Electronics Technologies, School of Mathematics and Physics, University of Queensland, St Lucia, Queensland 4072, Australia.}
\author{Matthew J. Davis}
\affiliation{Australian Research Council Centre of Excellence in Future Low-Energy Electronics Technologies, School of Mathematics and Physics, University of Queensland, St Lucia, Queensland 4072, Australia.}
\author{Matthew T. Reeves}
\affiliation{Australian Research Council Centre of Excellence in Future Low-Energy Electronics Technologies, School of Mathematics and Physics, University of Queensland, St Lucia, Queensland 4072, Australia.}
\date{\today}

\maketitle

This supplemental material is organised as follows.  We begin with an overview of the experimental procedure.  We then provide further details of the simulations, including the choice of parameters for the c-field simulations, how the current is calculated, and discuss the robustness of the results to varying the initial conditions and the energy cut-off of the c-field region. We then analyse the emergence of a condensate on the central site and examine the adiabaticity of the dynamics of the refilling of the system site.  Finally, we provide more results on the dependence of the dynamics on the initial coherence between the two sides of the Josephson chain, and describe simulations performed at finite temperature with a limited condensate fraction to demonstrate the absence of acceleration of filling without coherence.

\section{Review of the Experiment}
The experiment \cite{Labouvie2015} consists of an approximately 65 site 1D chain of 2D harmonic traps, see Fig.~\ref{fig:Schematic} of the main text, with a lattice spacing in the $z$-direction $a = 549$~nm and a radial trap-frequency $\omega_r= 165 \times 2\pi $ rad/s.  Traps near the centre of the chain hold approximately $N = 700$ atoms of Rb-87 with mass $m$ and an s-wave scattering length $a_s = 5.8$ nm. Sites near the ends of the array contain comparatively few sites due to the weak harmonic potential in the $z$-dimension. 

The experimental system was initialized as follows.  After an initial cooling process to produce the bulk condensate, a one-dimensional optical lattice is ramped up so that every site is effectively isolated in a deep well for 9~ms (note that $1 ~\text{ms} \approx  \omega_r^{-1}$). During this time the central site is also depleted via one-body losses provided by a site-selective electron beam. The final atom number on the central site is in the range of $5\%$ to $10\%$ of the full wells. The lattice height is then lowered to the final value over a time-scale of 2~ms, with the dissipation still turned on. 

The dissipation is then turned off and unitary dynamics ensues.  After the system evolves undisturbed for a period of time, electron impact ionization facilitates a measurement of the atom number. Since this process removes the atoms from the trap, only a single measurement can be made in a given experimental run. The entire procedure is then repeated for each time $t$ to obtain a statistically relevant sample. A detailed breakdown of the procedure can be found in Ref.~\cite{LabouvieThesis}.

\section{Details of simulations}
\subsection{Numerical method}
In this section we provide additional details of the c-field (classical field) numerical technique which we employ. For a broad overview of c-field methods see Ref.  \cite{Blakie2008}. The projection operator which defines the classical field region  is prescribed by an energy cutoff within the appropriate single-particle basis associated with Eq.~(\ref{eqn:GPEoperator}). We use a cartesian Hermite-Gauss computational basis, which diagonalizes the single-particle Hamiltonian associated with Eq.~(\ref{eqn:GPEoperator}) (i.e.  $g_2 = 0$); {this allows for the definition of the projector~Eq.~(\ref{eq:Projector}) in a suitable basis, and  provides an efficient computational basis, enabling the simulation of a large number of sites}. We define the Hermite-Gauss functions $\varphi_n(\bm{x})$ of the 2D harmonic oscillator, and the associated quantum numbers $n = n_x + n_y$. The cutoff $n_{\rm cut}$ is hence given via $\mathcal{C} = \{n \; | \; n_x + n_y \leq n_{\rm cut}\}$ and 
\begin{align}
\mathcal{P}\{f(\bm{x})\} = \sum_{n \in \mathcal{C}} \varphi_n(\bm{x}) \int d^2 \bm{x}' \varphi_n^*(\bm{x}') f(\bm{x}'). 
\label{eq:Projector}
\end{align}
Gaussian quadrature allows the matrix transforms to be evaluated exactly~\cite{Blakie2008,Boyd2001}. The number of retained modes (defined by $n_{\rm{cut}}$) is chosen to be sufficiently high that the relevant energy scales are captured, while ensuring that the results are not significantly affected by this choice. Specifically, we use $n_{\rm{cut}} = 3\mu/\hbar \omega_r$, where $\mu$ is the chemical potential of the ground-state.  This is chosen since collisions between excited atoms with energy $\mu$ that have tunnelled on to the depleted site can, occasionally, scatter to energies near $0$ and $2\mu$ respectively. Hence,  $2\mu$ is a potentially important scale. Accessing higher energy modes requires multiple scatterings of energetic atoms. Some of these processes are captured by using a cut-off of  $3\mu$. 

\begin{figure}[t]
\includegraphics[width=\columnwidth]{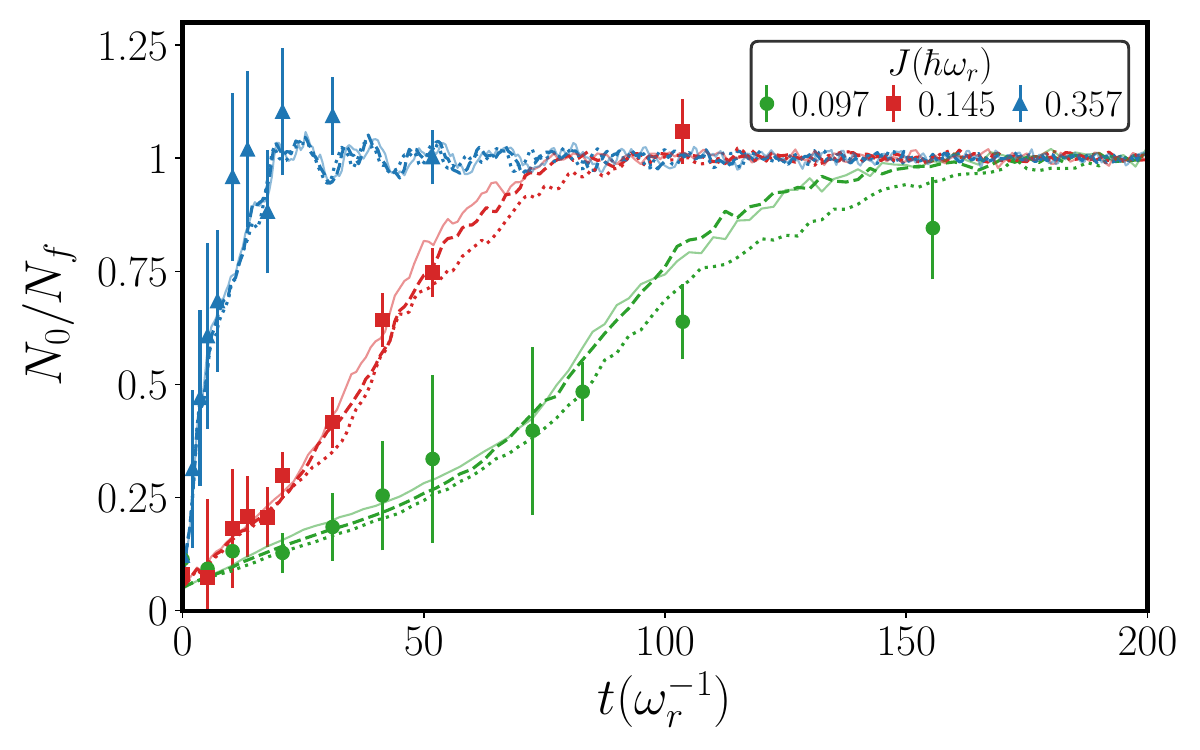}
\caption{Atom number $N_0/N_f$ vs time for different tunnel couplings $J$, comparing results obtained via  the c-field model (\ref{eq:gpe}) with the experimental data of Ref.~\cite{Labouvie2015} (points). The dashed lines correspond to the c-field model with a mode cut-off $n_{cut} = 3\mu/\hbar \omega_r$ (used throughout the manuscript),  while the dotted lines are for $n_{cut} = 2\mu/\hbar \omega_r$ and light solid lines are for $n_{cut} = 4\mu/\hbar \omega_r$. We use 100 trajectories for the lower cutoffs, but only 40 for the case of  $n_{cut} = 4\mu/\hbar \omega_r$ due to the increased computation time.}
\label{fig:supp_fullns}
\end{figure} 

Figure~\ref{fig:supp_fullns} shows the filling for cut-offs of $n_{\rm{cut}} = \{2\mu/\hbar \omega_r , 3\mu/\hbar \omega_r ,4\mu/\hbar \omega_r \}$. Due to the 2D nature of each site, $4\mu/\hbar \omega_r$  corresponds to approximately $3.5$ times the number of modes as  $2\mu/\hbar \omega_r$. 
The results suggest that the lower cut-off $n_{\rm{cut}} = 2\mu/\hbar \omega_r$ only changes the results slightly from the larger cut-offs, primarily for the case of $J = 0.097 \hbar \omega_r$. We suspect that at low $J$ values the slower dynamics simply allow more time for successive scattering events that can populate the high energy modes, and therefore more modes are required. The results for cut-offs of  $3\mu$ and $4\mu$ are almost identical, suggesting they are sufficiently high to capture the scales of interest. Discrepancies between these results are of the order of the sampling error for the $4\mu$ case (not shown).

\subsection{Simulation parameters} \label{sec:param}
We now discuss how the model parameters were determined. Using the details provided in the previous section, the 3D interaction strength can be calculated as $g = 4\pi\hbar^2 a_s/m$. Since the model Eq.~(\ref{eq:gpe}) considers the $z$-direction to be discrete due to the strong confining potential, we obtain the 2D interaction strength $g_2 = g \int dz ~ w(z)^4 ,$ where $w(z)$ is the Wannier band in the $z$-direction, which is obtained numerically. The chemical potential is obtained via the Thomas-Fermi approximation: $\mu =\sqrt{g_2 m \omega_r^2 N /\pi}$. 

 We use the specific values of $J$ provided in Ref.~\cite{Labouvie2015}, which can be calculated from first principles by considering the corresponding lattice potential $V_0 \sin^2(k z)$ in the $z$-direction. We determine $V_0$ using a harmonic oscillator approximation in the center of the well: $V_0 = \hbar^2 \omega_z^2/4 E_r$, where $E_r = \hbar^2 k^2/2m$ is the lattice recoil energy, and we define the lattice wavevector as $k= 2 \pi/\lambda$, with lattice spacing $a$ and lattice wavelength $\lambda = 2a$.
 Using the tight-binding approximation, the tunneling can be obtained as $J = \Delta E/4$ where $\Delta E$ is the band-width of the lowest Bloch band \cite{Arzamasovs2017}. We obtain $\Delta E$ by diagonalizing $\hat{H} = - \hslash^2 \partial_z^2/2m + V_0 \sin^2(2 \pi z / \lambda)$ numerically. Since $J$ is in one-to-one correspondence with $\omega_z$, we check that fixing $J$ produces $\omega_z$ values in the range proposed by \cite{Labouvie2015}. We further check that the results do not deviate significantly from the exact expression obtained using Matheiu functions in the `deep lattice' limit $V_0 \gg E_r$  \cite{Zwerger2003}.  The above procedure uniquely determines $\omega_z$, $g_2$, and $\mu$, which are required for simulations. 

\subsection{Initial conditions}
While the experiment Ref.~\cite{Labouvie2015} contained 65 sites, for computational expediency, we consider a somewhat smaller but comparable uniform system of 21 sites. As the total atom number is conserved by Eq.~(\ref{eq:gpe}), the atom number in equilibrium is thus reduced by $\sim5\%$ across each site to refill the central site (cf.~$\sim$1--2\% for the 65-site chain). For the tunneling strengths $J$ relevant to the experiment (determined by lattice height $V_0$), the effective 2D interaction strength is $g_2 \sim 0.2 \hbar\omega_r l_r^2$, and the chemical potential of an isolated site is $\mu \sim 7 \hbar \omega_r$ relative to the bottom of the central site. Note these values vary slightly as $J$ is varied (via $V_0$).

To initialize simulations, we first find the system's zero temperature ground-state by evolving  Eq.~(\ref{eq:gpe}) in imaginary time. We then remove almost all atoms on the central site ($i=0$), such that the initial relative atom number is $\sim5\%$, consistent with the experimentally reported values. 

The removal of the atoms at the central site effectively separates the half-chains on each side into two isolated systems; the relative phase between left and right chains are thus expected to drift due to quantum (or  thermal) fluctuations~\cite{Sinatra2008} while the two subchains are disconnected.  Figure~1 shows how the relative phase  evolves due to quantum fluctuations; for the parameters of the experiment.  The phase difference is shown to grow over a time of $30 \sim 40 ~~ \omega_r^{-1}$, after which point it has reached approximately $\pi/2$, indicating complete randomness. In the experiment, the half-chains are uncoupled for 11 ms.  Our results suggest this is long enough for significant phase randomness to develop between the half-chains. For small values of $J$ the initial part of the filling process is quite slow, and the dephasing should continue even after the dissipation is turned off.  Furthermore, thermal fluctuations will further contribute to the decoherence.  Considering these factors, we conclude that the phase of each half-chain in the experiment may be modelled, to a good degree of approximation, as independent random variables. Thus, for the simulations presented in the main text,  for each trajectory, we therefore multiply the right $R = \{i | i > 0\}$ subset of the chain by random phase $e^{i \theta}$, $\theta \in [0,2\pi]$, to mimic this phase diffusion. 
Finally, half a particle per mode of noise is added to all sites, to capture vacuum fluctuations in accordance with the TWA~\cite{Steel1998,Polkovnikov2003a,Blakie2008}.

 \begin{figure}[t]
     \centering
     \includegraphics[width =\columnwidth]{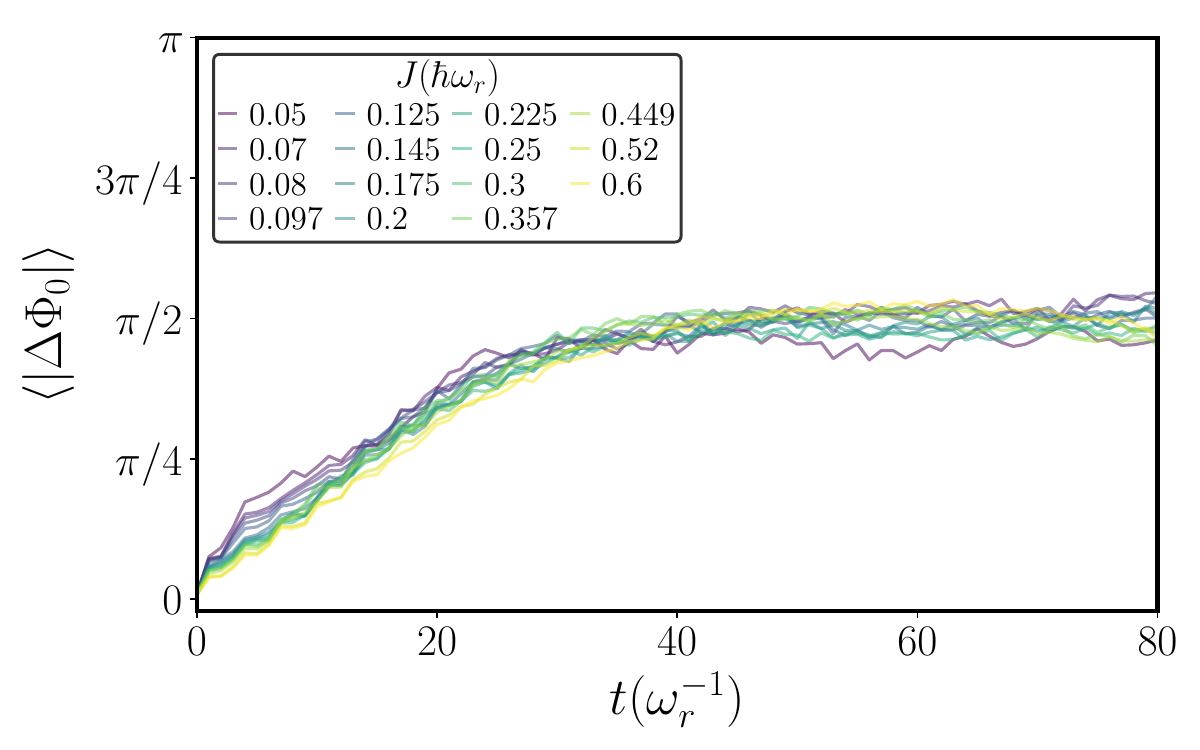}
    \caption{Time evolution of the mean phase difference, $\langle |\Delta \Phi_0| \rangle$, between two disconnected chains containing 10 sites. The chains are initially identical up to quantum fluctuations and close to the zero temperature ground-state (see discussion in text). The results correspond to an average over 100 independent trajectories for each value of $J$. After a time-scale of approximately $30 \omega_r^{-1}$ the phase difference has approached the value of $\pi/2$, indicating complete randomness.}
     \label{fig:halfpahse_ev}
 \end{figure}

\subsection{Dependence on the initial atom number}
\begin{figure}[t]
\includegraphics[width=\columnwidth]{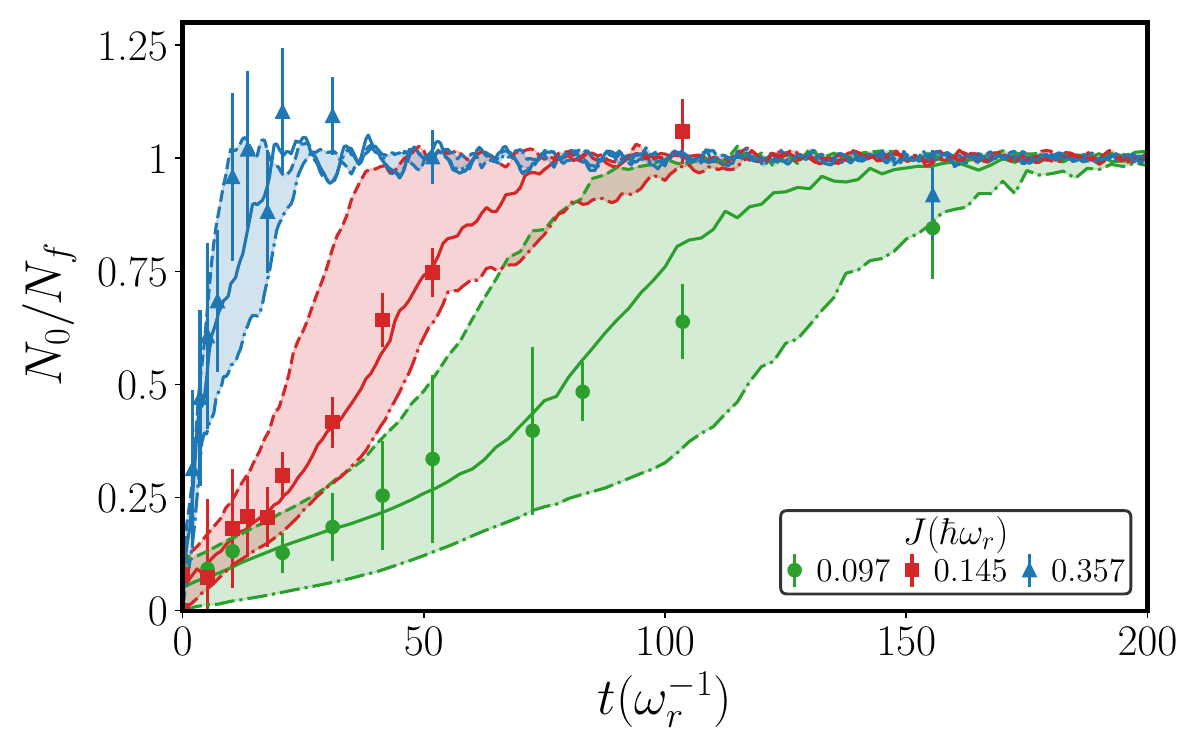}
\caption{Comparison of filling curves for three initial conditions with differing atom number on the central site. The experimental data (dots) is compared against an initial population of 10\% (dotted lines), 5\% (solid lines) and 0\% (empty well, dashed lines). The shaded regions,  bounded by the  0\% and 10\% cases, illustrate the range of results.  }
\label{fig:diffinit}
\end{figure}

We briefly discuss the dependence of the filling dynamics on the initial population of the central site. We find that the results are insensitive to moderate changes in the initial atom number. This is demonstrated in Fig.~\ref{fig:diffinit}, in which we compare the atom number vs time for the cases in which the initial population is 0\%, 5\% and 10\% of a full well. The 5\% and 10\% cases correspond to the limits of the experimental values reported in \cite{Labouvie2015}. It can be seen that the filling curves follow broadly the same trend, with the 10\% case giving faster filling. This is anticipated from the increased Franck-Condon factor Eq.~(\ref{eqn:FrankCondon}) in this case.

\subsection{Current calculation}
The current $I_t \equiv {dN_0(t)}/{dt}$ for an individual trajectory is calculated directly from Eq.~(4) of the main text at each time $t$.
At a given time, we also calculate $\Delta N(t) = (N_f-N_0(t))/N_f$ for each trajectory, which gives  us the ``current-voltage'' (current-atom number difference) pair $(dN_0(t)/dt,\Delta N(t))$. To calculate the statistics of these quantities as shown in Fig.~\ref{fig:mean_vs_trajectories}(c), the data for every time-step is placed in narrow bins according to the value of $\Delta N(t)$. This process is repeated for every trajectory. The average and fluctuations can then be obtained from the statistics within a given bin.

In contrast, the quantity we call $I_m$ is calculated from the mean (trajectory-averaged) atom number $\langle N_0(t)\rangle$ at each time $t$, i.e. $I_m = d\langle N_0(t)\rangle/dt = [{\langle N_0(t)\rangle - \langle N_0(t-\delta) \rangle}]/{\delta}.$  This is how the current was inferred in Ref. \cite{Labouvie2015}, since only average values are available due to the destructive measurement protocol.

 \begin{figure}[t]
 \centering
 \subfloat[]{\includegraphics[width=8.7cm]{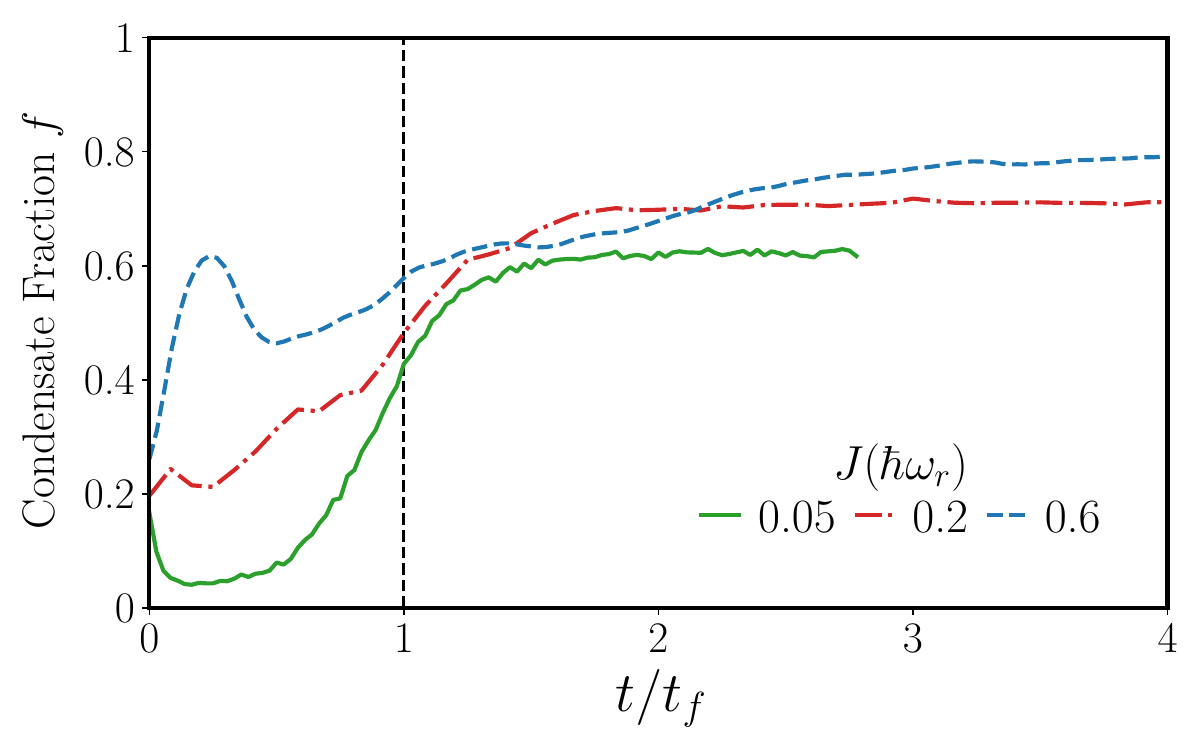}\label{fig:condfrac}}

 \vspace{-0.3cm}
 \centering

 \subfloat[]{  \hspace{0.5cm}\includegraphics[width=8.0cm]{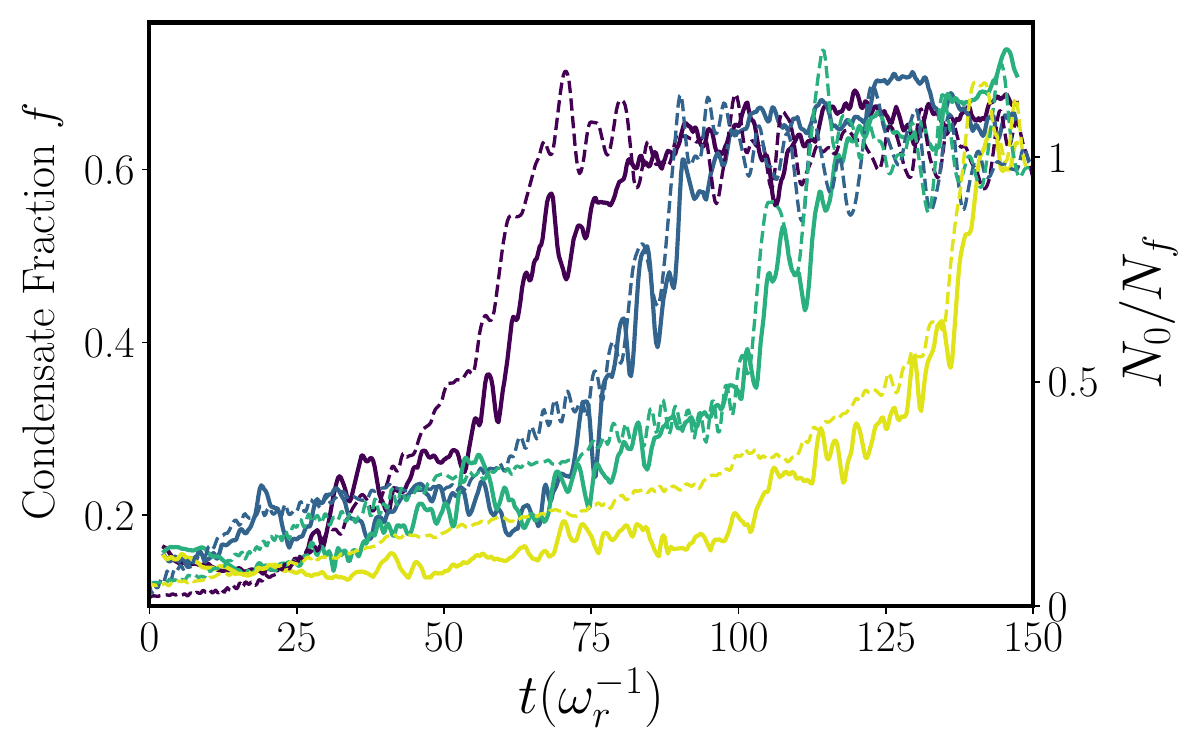}\label{fig:condtraj}}
 \caption{(a) Condensate fraction $f$ vs time for different tunnel couplings $J$. (b) Condensate fraction for four individual trajectories in the case of $J = 0.097\hbar\omega_r$ (solid lines, left axis values).  The atom number $N_0/N_f$ vs time is also shown (dashed lines, right axis values), with matching line colors for each trajectory. These are the same trajectories as those shown in Fig.~\ref{fig:mean_vs_trajectories}(a) of the main text.  For each trajectory, the condensate fraction at each time is calculated from the state at 20 times over a window of $ 5 ~\omega_r^{-1}$ centered on the given time. }
 \label{fig:cond_things}
 \end{figure}
 \section{Dynamics of condensation}
 In this section we consider the dynamics of the condensate fraction on the central site.  The condensate fraction can be determined in c-field simulations by calculating the single-particle density matrix $\rho_{ij} = \langle \psi_i^*|\psi_j\rangle$, and then diagonalizing it.  The largest eigenvalue is then the condensate number.  The average can either be calculated over trajectories in non-equilibrium scenarios, or as a time-average from a single trajectory in a quasi-equilibrium situation.  See Ref.~\cite{Blakie2008} for further details.
 
 Figure~\ref{fig:cond_things}(a) shows the time evolution of the condensate fraction $f$ for a variety of coupling values, which match to those displayed in Fig.~\ref{fig:relativePhase} of the main text. We observe that for weak couplings ($J = 0.05 \hbar \omega_r$) the condensate fraction drops to  very small values before rising as the filling time is approached. The incoherent behavior at early times explains the observations in Fig.~\ref{fig:relativePhase}(a) of the main text, which demonstrated independence of the filling time on the initial phase difference across the central site. Although outside the scope of the present work, the peak at early times for $J = 0.6 \hbar\omega_r$ is due to the trajectories for which $|\Delta \Phi_0|$ is initially small. The slow rise that occurs after this is due to the trajectories for which $|\Delta \Phi_0|$ is initially larger, 
   such that the neighboring wells are out of phase. This is related to the dark solitons discussed in Ref.~\cite{Ceulemans2023} and will be reported on elsewhere.

 Figure \ref{fig:cond_things}(b) shows an estimate of the short-time averaged condensate fraction for four individual trajectories in the case of  $J = 0.097 \hbar\omega_r$. The single-particle density matrix
 at a given time is calculated by time-averaging the single-particle density matrix over a rolling time-window of $5~\omega_r^{-1}$ using 20 samples of $\psi_i$.
 We choose the same trajectories as in Fig.~\ref{fig:mean_vs_trajectories}(a) with matching colors for the lines. A comparison of the data against the atom number $N_0/N_f$ for each trajectory (dashed lines) shows that the onset of rapid growth in the atom number is highly correlated with the onset of significant condensation.

\begin{figure}[t]
\subfloat{\includegraphics[width=\columnwidth]{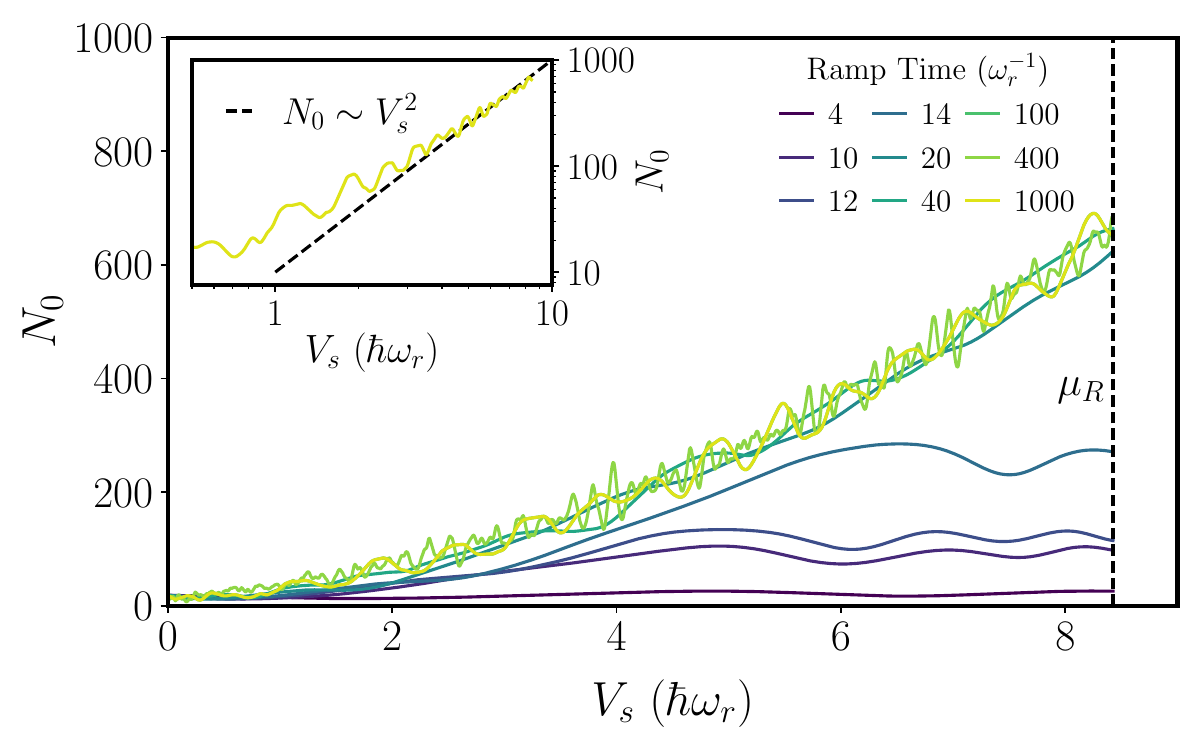}}
\caption{Atom number $N_0$ vs energy shift $V_s(t)$  for the ramping procedure defined by Eq.~(\ref{eq:gperamp}) and Eq.~(\ref{eq:ramp}), comparing a variety of ramp times (legend) for weak tunnel coupling $J/\hbar\omega_r = 0.05.$
The inset compares the slowest ramp, conducted over an interval of $1000 \omega_r^{-1}$, against the Thomas-Fermi prediction of $N_0 \sim V_s^2 \equiv \mu_s^2$, plotted on a log-log scale. }
\label{fig:ramp}
\end{figure}

\section{Far-from-equilibrium nature of system dynamics}
In this section we demonstrate that in the experiment of Labouvie~\emph{et al.}~\cite{Labouvie2015} the system is in fact far-from-equilibrium during the filling process, and hence cannot be considered to have an effective chemical potential. To do this we perform a numerical experiment in which we initially add an energy shift of $(\mu_R - \hbar\omega_r)$ to the central depleted site.  This means that that the  chemical potential of the undepleted sites is equal to the vacuum energy of the two-dimensional harmonic trap on the central site, and hence it remains unfilled in equilibrium.
 
Beginning from this equilibrium state with the central site empty and the other sites full, we then  decrease the energy shift on the central site to zero linearly in time at a rate $\nu$. This introduces a  potential difference between the central site and the rest of the lattice that results in atoms tunnelling onto the central site and refilling it. When the energy shift is eventually removed, the central site will fill until it  contains the same number of atoms as on the sites of the rest of the lattice. 

The relevant  coupled Gross-Pitaevskii equation for the system site is given by
\begin{align}
i \hbar \partial_t \psi_0 = \mathcal{L}\psi_0 + J (\psi_{-1}  + \psi_{1}) + \Delta V(t) \psi_0 , \label{eq:gperamp}
\end{align}
with the energy shift for $0 \le \nu t \le 1$ being 
\begin{eqnarray}
\Delta V(t)  &=& (1-\nu t)(\mu_R -\hbar \omega_r),\nonumber\\
&=& (\mu_R -\hbar \omega_r) - V_s(t), \label{eq:ramp}
\end{eqnarray}
where $\nu$ is the ramp rate, $V_s(t) = \nu t(\mu_R -\hbar \omega_r)$, and the equation for other sites 
is unchanged
as in Eq.~(\ref{eq:gpe}).

\begin{figure*}[t]
\subfloat{\includegraphics[width=2.0\columnwidth]{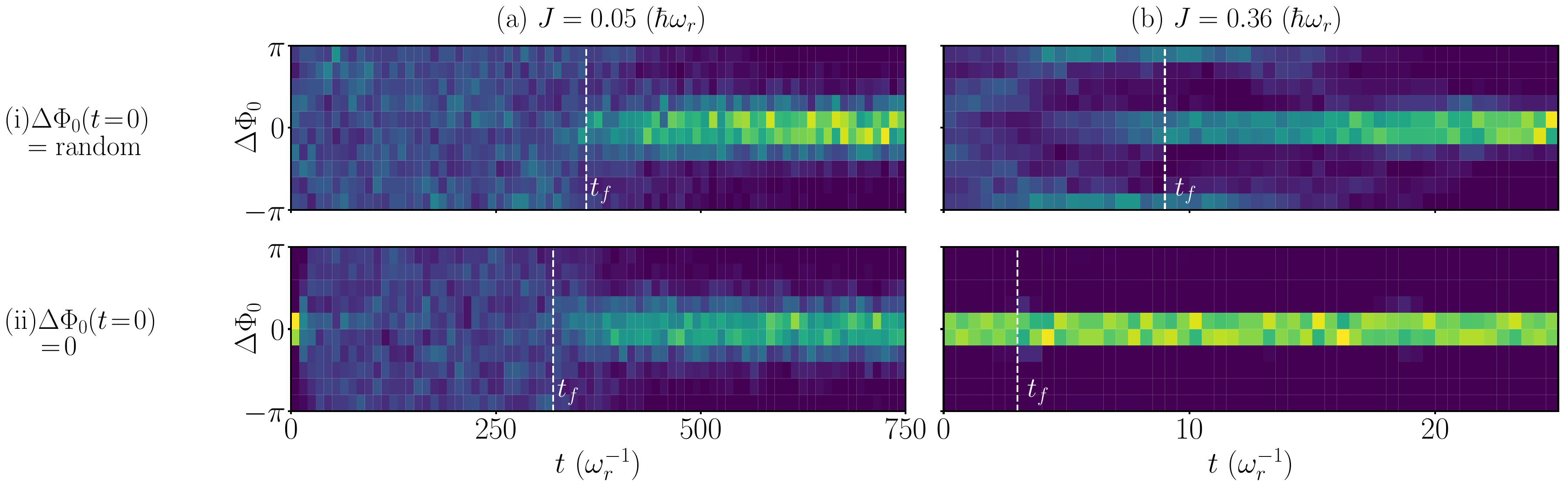}\label{fig:ph_time_colorplot}} 
\vspace{-0.0cm}

\hspace{3.0cm} \includegraphics[width=0.37\columnwidth]{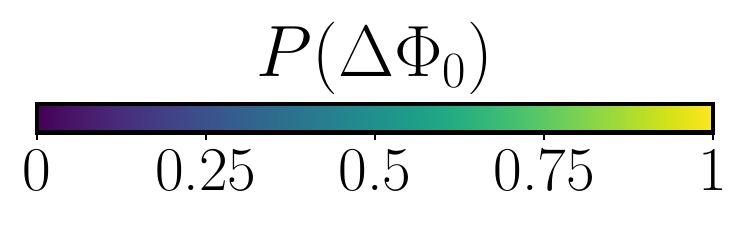}
\caption{Probability distribution of the relative phase $\Delta \Phi_0$ vs. time, for an initially random phase difference between the left and right chains (i), and no phase difference (ii). The left column (a) corresponds to $J = 0.05 \hbar \omega_r$, while the right column (b) is $J = 0.36 \hbar \omega_r$  The vertical lines indicate the average filling time $t_f$ of the trajectories.}
\label{fig:phasehist}
\end{figure*}

If this ramp is conducted slowly enough, the 
system will remain in quasi-equilibrium, and the central site will have the same occupation for a given value of $V_s(t)$ regardless of the ramp rate.
Figure~\ref{fig:ramp} shows the atom number $N_0$ vs the energy $V_s(t)$ for different ramp times $1/\nu$, with $J/\hbar\omega_r = 0.05$. For slow ramps, conducted over a time of roughly $20~\omega_r^{-1}$ or more, we observe that all ramps result in essentially the same $N_0$  vs $V_s$ curve, showing that the system is in a quasi-equilibrium state. These scenarios allow for the definition of an on-site chemical potential $\mu_S(t) = V_s(t) + \hbar \omega_r$. Provided the atom number $N_0$ exceeds 200 or so, the Thomas-Fermi prediction of $N_0 \sim (\Delta \mu)^2 $ is a reasonable approximation, as shown by the dotted line in the inset. 

For ramp times shorter than $20~\omega_r^{-1}$, the system no longer follows the same curve and hence the dynamics are non-equilibrium, and a chemical potential for the system cannot be defined.  In the experiment of Labouvie~\emph{et al.}~\cite{Labouvie2015}, the rapid change of the lattice height to initialise the refilling dynamics occurs over a timescale of approximately $2$~ms or $\sim 2 \omega_r^{-1}$ during the preparation period.  The results presented here clearly show that the system is  far-from-equilibrium.

 \section{Phase Dynamics}

 In Fig.~4 of the main text, we showed the system filling dynamics is dependent on the initial phase difference across the central well, $\Delta \Phi_0$, at high $J$ but not at low $J$. This section provides further discussion on this dependence. Further insight into the role of the phase difference is gained by inspecting the statistics of $\Delta \Phi_0$ across different trajectories; to this end, in Fig.~\ref{fig:ph_time_colorplot}, we show the probability distribution $P(\Delta \Phi_0)$, for small (a) and large (b) coupling strengths $J$.  The top panels (i) show the situation in which the left and right chains are initialized randomly, while the bottom panels (ii)  show the case where the chains are in phase ($\Delta \Phi_0=0 $).

 In Fig.~\ref{fig:phasehist}(a),  (small $J$) [Fig.~\ref{fig:phasehist}(a)], it does not matter whether the relative phase between the left and right chains is initially random ($\Delta \Phi_0 \in [0,2\pi]$ uniformly, top row) or synchronized ($\Delta \Phi_0 =0$, bottom row). The initial phase is quickly forgotten, and rephasing occurs at a similar time in both cases, $t_f \omega_r \sim 320-350$, which is within the sampling error. By contrast, for large $J$ [Fig.~\ref{fig:phasehist}(b)] the initial phase is crucial to the refilling dynamics. For random phases (top) some trajectories gravitate towards $\Delta \Phi = \pm \pi$ and remain there for some time, dramatically slowing the filling. This is related to the formation of dark solitons \cite{Ceulemans2023}, and will be discussed elsewhere. Meanwhile, for synchronized phases (bottom), the filling is comparatively rapid.

\begin{figure*}[t]
\includegraphics[width = 2\columnwidth]{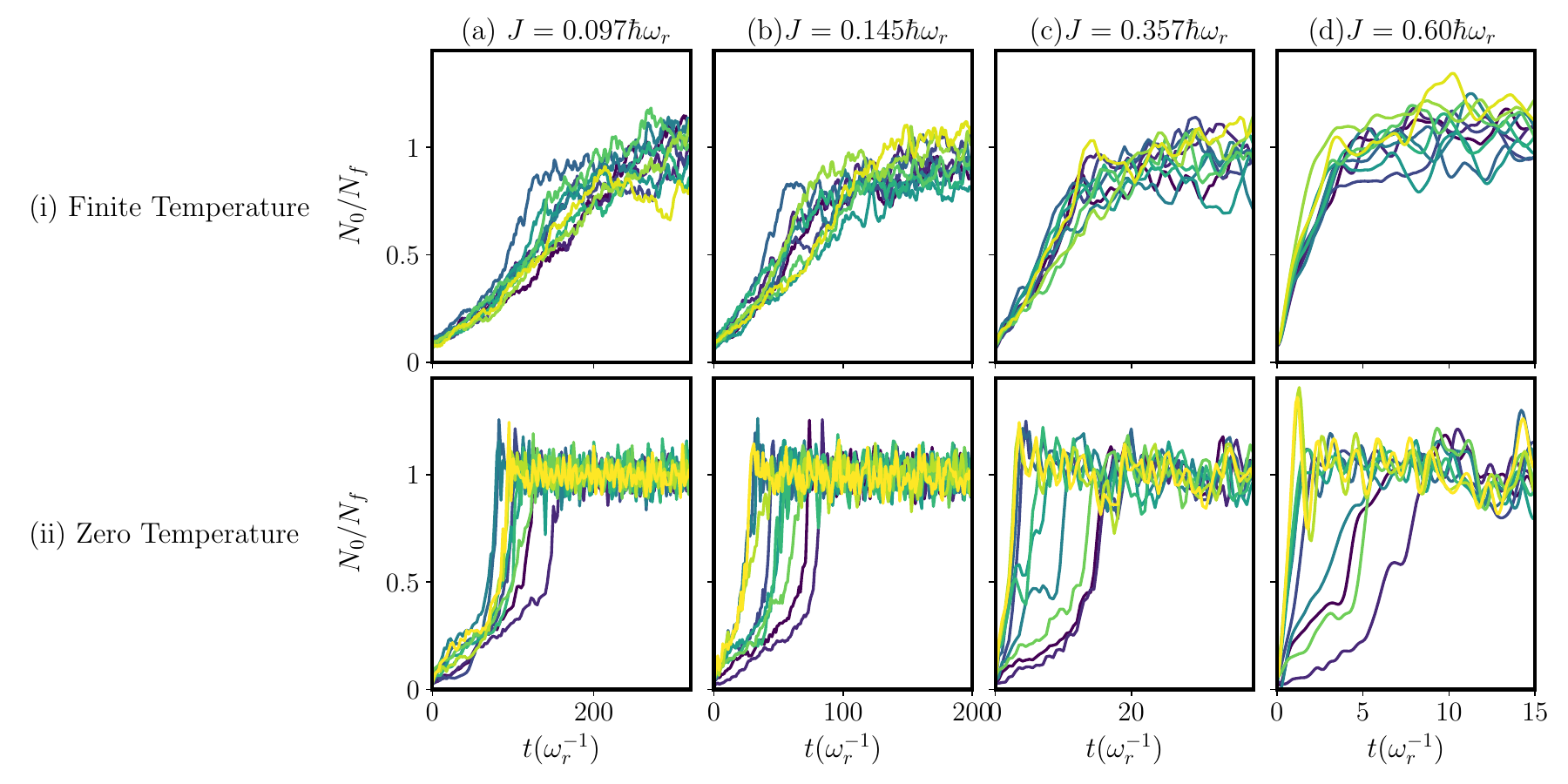}
\caption{Top line (i): Atom number $N_0/N_f$ vs time for 10 individual trajectories (randomly chosen) at finite temperature and different values of the tunnel coupling (a)-(d). Bottom line (ii): same but at zero temperature. Note that the data is saved at larger intervals (relative to $t_f$) than in Fig.~\ref{fig:mean_vs_trajectories}(a) of the main text, so some high frequency oscillations may not be shown.}
\label{fig:zerotemp}
\end{figure*}

 \section{Finite Temperature Case}
In the main text, we argue that the acceleration of the current seen in trajectories is due to the onset of condensation and phase coherence. To provide further evidence of this interpretation, in this section we provide additional simulations of the system but at finite temperature (i.e., low initial condensate fraction on the filled sites).

To prepare a finite temperature initial state, we unitarily evolve an arbitrary initial state with $E/N_f = 10 \hbar \omega_r,$ allowing it to thermalize to the equilibrium state with this energy. We ensure that the atom number is comparable to the atom number in the  experiment. The initial condensate fraction on the filled sites for this scenario is less than $5$\%. The results for some individual trajectories are shown in Fig.~\ref{fig:zerotemp}(i), for a variety of $J$ values. For comparison, Fig.~\ref{fig:zerotemp}(ii) shows data for the zero temperature cases considered in the main text. In the finite temperature case, the vast majority of the trajectories do not show an  acceleration in the filling rate during the later stages of the filling dynamics. Rather, in this case the current often slows down over time. This supports our interpretation of this process being driven by condensation and phase coherence. There is some evidence of an NDC characteristic for low $J$ and early times, which would align with the `incoherent' filling interpretation as was put forward in Ref.~\cite{Labouvie2015}. However, this is not a prominent feature compared to some individual trajectories for the zero temperature cases in Fig.~\ref{fig:zerotemp}(ii). Interestingly, a direct comparison of these figures suggests that the fluctuations are actually significantly reduced at finite temperature. This may be due to the lack of Josephson oscillations at high temperatures, as these cause large fluctuations in the coherent case.

Further argument for our interpretation in terms of coherence can be drawn from the spatial overlap integral in Eq.~(\ref{eqn:FrankCondon}) of the main text; this equation shows that the current into site $i$ is  $\propto \sin(\varphi_{ij})$, the relative phase difference between neighbouring sites. When the depleted site has no well-defined phase, this term averages to a value near zero. However, once condensation occurs, the depleted site adopts a definite phase, allowing this term to be $\sim \mathcal{O}(1)$.   This leads to a larger current in Eq.~(\ref{eq:fillrate}), which would appear to explain the late time acceleration in the filling observed in Fig.~\ref{fig:mean_vs_trajectories}(a) of the main text. However, since the overlap is complex this does not rule out Josephson oscillations; it is also necessary that atoms tunneling into the central site can be irreversibly scattered into a large number of modes. This is guaranteed since the condensate fraction, while finite and growing, is still far from unity as shown in Fig.~\ref{fig:cond_things}.
The effective driving frequency is also changing with time, which may cause some damping.

%\bibliography{References.bib}
\bibliographystyle{apsrev4-2}

\clearpage